*Review Article*

# Semantic Web in Healthcare: A Systematic Literature Review of Application, Research Gap, and Future Research Avenues

**A. K. M. Bahalul Haque**,[1] **B. M. Arifuzzaman**,[1] **Sayed Abu Noman Siddik**,[1] **Abul Kalam**,[1] **Tabassum Sadia Shahjahan**,[1] **T. S. Saleena**,[2] **Morshed Alam**,[3] **Md. Rabiul Islam**,[4] **Foyez Ahmmed**,[5] **and Md. Jamal Hossain**[6]

[1]*Electrical and Computer Engineering, North South University, Dhaka 1229, Bangladesh*
[2]*PG & Research Department of Computer Science, Sullamussalam Science College Areekode, Malappuram, Kerala 673639, India*
[3]*Institute of Education and Research, Jagannath University, Dhaka 1100, Bangladesh*
[4]*Department of Pharmacy, University of Asia Pacific, 74/A Green Road, Farmgate, Dhaka 1205, Bangladesh*
[5]*Department of Statistics, Comilla University, Kotbari, Cumilla, Bangladesh*
[6]*Department of Pharmacy, State University of Bangladesh, 77 Satmasjid Road, Dhanmondi, Dhaka 1205, Bangladesh*

Correspondence should be addressed to A. K. M. Bahalul Haque; bahalul.haque@lut.fi and Md. Jamal Hossain; jamalhossain@sub.edu.bd

Today, healthcare has become one of the largest and most fast-paced industries due to the rapid development of digital healthcare technologies. The fundamental thing to enhance healthcare services is communicating and linking massive volumes of available healthcare data. However, the key challenge in reaching this ambitious goal is letting the information exchange across heterogeneous sources and methods as well as establishing efficient tools and techniques. Semantic Web (SW) technology can help to tackle these problems. They can enhance knowledge exchange, information management, data interoperability, and decision support in healthcare systems. They can also be utilized to create various e-healthcare systems that aid medical practitioners in making decisions and provide patients with crucial medical information and automated hospital services. This systematic literature review (SLR) on SW in healthcare systems aims to assess and critique previous findings while adhering to appropriate research procedures. We looked at 65 papers and came up with five themes: e-service, disease, information management, frontier technology, and regulatory conditions. In each thematic research area, we presented the contributions of previous literature. We emphasized the topic by responding to five specific research questions. We have finished the SLR study by identifying research gaps and establishing future research goals that will help to minimize the difficulty of adopting SW in healthcare systems and provide new approaches for SW-based medical systems' progress.

## 1. Introduction

The detection and remedy of illnesses through medical professionals are expressed as healthcare. The healthcare system consists of medical practitioners, researchers, and technologists that work together to provide affordable and quality healthcare services. They tend to generate considerable amounts of data from heterogeneous sources to enhance diagnostic accuracy, elevate quick treatment decisions, and pave the way for the effective distribution of information between medical practitioners and patients. However, it is necessary to organize those valuable data appropriately so that they can fetch those, while required.

One of the main challenges in utilizing medical healthcare data is extracting knowledge from heterogeneous data sources. The interoperability of well-being and clinical information poses tremendous obstacles due to data irregularity and inconsistency in structure and organization [1, 2]. This is also because data are stored in various authoritative areas, making it challenging to retrieve knowledge and authorize a primary route along with information analysis. The information from a hospital can prove to be very useful in healthcare if these data are shared, analyzed, integrated, and managed regularly. Again, platforms that provide healthcare services also face dilemmas in automating time-efficient and low-cost web service arrangements [3]. This indicates that meaningful healthcare solutions must be proposed and implemented to provide extensive functionality based on electronic health record (EHR) workflows and data flow to enable scalable and interoperable systems [4], such as a blockchain-based smart e-health system that provides patients with an easy-to-access electronic health record system through a distributed ledger containing records of all occurrences [5–8]. A standard-based and scalable semantic interoperability framework is required to integrate patient care and clinical research domains [9]. The increasing number of knowledge grounds, heterogeneity of schema representation, and lack of conceptual description make the processing of these knowledge bases complicated. Non-experts find mixing knowledge with patient databases challenging to facilitate data sharing [10]. Similarly, ensuring the certainty of disease diagnosis also becomes a more significant challenge for health providers. Brashers et al. [11] in their work examined the significance of credible authority and the level of confidence HIV patients have in their medical professionals. Many participants agreed that doctors might not be fully informed of their ailment, but they emphasized the value of a strong patient-physician bond. With the help of big data management techniques, these challenges can be minimized. Likewise, Crowd HEALTH aims to establish a new paradigm of holistic health records (HHRs) that incorporate all factors defining health status by facilitating individual illness prevention and health promotion through the provision of collective knowledge and intelligence [11–13]. Another similar approach is adopted by the beHealthier platform which constructs health policies out of collective knowledge by using a newly proposed type of electronic health records (i.e., eXtended Health Records (XHRs)) and analysis of ingested healthcare data [14]. Making healthcare decisions during the diagnosis of a disease is a complex undertaking. Clinicians combine their subjectivity with experimental and research artifacts to make diagnostic decisions [9].

In recent years, Web 2.0 technologies have significantly changed the healthcare domain. However, in proportion to the growing trend of being able to access data from anywhere, which is primarily driven by the widespread use of smartphones, computers, and cloud applications, it is no longer sufficient. To address such challenges, *Semantic Web Technologies* have been adopted over time to facilitate efficient sharing of medical knowledge and establish a unified healthcare system. Tim Berners-Lee, also known as the father of the web, first introduced Semantic Web (SW) in 1989 [15]. The term "Semantic Web" refers to linked data formed by combining information with intelligent content. SW is an extension of the World Wide Web (WWW) and provides technologies for human agents and machines to understand web page contents, metadata, and other information objects. It also provides a framework for any kind of content, such as web pages, text documents, videos, speech files, and so on. The linked data comprise technologies such as Resource Description Framework (RDF), Web Ontology Language (OWL), SPARQL, and SKOS. It aims to create an intelligent, flexible, and personalized environment that influences various sectors and professions, including the healthcare system.

Data interoperability can only be improved when the semantics of the content are well defined across heterogeneous data sources. Ontology is one of the semantic tools, which is frequently used to support interoperability and communication between software, communities, and healthcare organizations [16, 17]. It is also commonly used to personalize a patient's environment. Kumari et al. [18] and Haque et al. [19] proposed an Android-based personalized healthcare monitoring and appointment application that considers the health parameters such as body temperature, blood pressure, and so on to keep track of the patient's health and provide in-home medical services. Some existing ontologies of medicine are Gene, NCI, GALEN,-LinkBase, and UMLS [20]. They have also been used in offering e-healthcare systems based on GPS tracking and user queries. Osama et al. proposed two ontologies for a medical differential diagnosis: disease symptom ontology (DSO) and patient ontology (PO) [21]. Sreekanth et al. used semantic interoperability to propose an application that brings together different actors in the health insurance sector [22]. Semantic Web not only enables information system interoperability but also addresses some of the most challenging issues with automated healthcare web service settings. SW combined with AI, IoT, and other technologies has produced a smart healthcare system that enables the standardization and depiction of medical data [1, 23, 24]. In terms of economic efficiency, the Semantic Web-Based Healthcare Framework (SWBHF) is said to benchmark the existing BioMedLib Search Engine [25]. SW also offered a new user-oriented dataset information resource (DIR) to boost dataset knowledge and health informatics [26]. This technology is also used in the rigorous registration process to discover, classify, and composite web services for the service owner [4]. To provide answers to medical questions, it has been integrated with NLP to create RDF datasets and attach them with source text [27]. Babylon Health, which enables doctors to prescribe medications to patients using mobile applications, has benefited from the spread of semantic technology. Archetypes, ontology, and datasets have been used in web-based methods for diagnosing colorectal cancer screening. Clinical information and knowledge about disease diagnosis are encoded for decision making with the use of ontological understanding and probabilistic reasoning. The integration of pharmaceutical and medical knowledge, as well as IoT-enabled smart cities, has made extensive use of SW technologies [8]. To put it briefly, this emerging technology has revolutionized the healthcare and medical system.

Despite its relevance, researchers who looked into the benefits of SW efforts showed substantial deficiencies in the wide range of semantic information in the medical and healthcare sectors. To the best of our knowledge, no previous systematic literature review (SLR) has been published on the Semantic Web and none of the research has previously classified the precise application area in which SW can be applied. Furthermore, there was an absence of research questions in the previous literature for analyzing and comparing similar works in order to understand their flaws, strengths, and problems.

In this study, we present a systematic review of the literature on Semantic Web in healthcare, with an emphasis on its application domain. It is absolutely essential to point the SW user community in the right direction for future research, to broaden knowledge on research topics, and to determine which domains of study are essential and must be performed. Thus, the current SLR can help researchers by addressing a number of factors that either limit or encourage medical and healthcare industries to employ Semantic Web technologies. Furthermore, the study also identifies various gaps in the existing literature and suggests future research directions to help resolve them. The research questions (RQs) that this systematic review will seek to answer are as follows. (**RQ1**) What is the research profile of existing literature on the Semantic Web in the healthcare context? (**RQ2**) What are the primary objectives of using the Semantic Web, and what are the major areas of medical and healthcare where Semantic Web technologies are adopted? (**RQ3**) Which Semantic Web technologies are used in the literature, and what are the familiar technologies considered by each solution? (**RQ4**) What are the evaluating procedures used to assess the efficiency of each solution? (**RQ5**) What are the research gaps and limitations of the prior literature, and what future research avenues can be derived to advance Web 3.0 or Semantic Web technology in medical and healthcare?

This research contributes in a number of ways. This paper's main focus is centered on the collection of some statistical data and analysis results that are mostly focused on the adoption of SW technologies in the medical and healthcare fields. First, we gathered data from five publishers, including Scopus, IEEE Xplore Digital Library, ACM Digital Library, and Semantic Scholar, to thoroughly review, analyze, and synthesize past research findings. Furthermore, the current study does not focus on a specific theme, rather, it offers a broad overview of all possible research themes related to the use of SW in healthcare. Finally, this SLR identifies gaps in the existing literature and suggests a future research agenda. The primary contributions of our study are listed as follows:

(i) To find out the up-to-date research progress of SW technology in medical and healthcare.
(ii) To open up new technical fields in healthcare where SW technologies can be used.
(iii) To identify all the constraints in the healthcare industry during the adoption of SW technologies.
(iv) To identify key future trends for semantics in the healthcare sector.
(v) To analyze and investigate alternative strategies for ensuring semantic interoperability in the healthcare contexts.

This review paper is organized as follows. Section 1 introduces Semantic Web technologies in healthcare followed by Section 2 which describes the methodology followed, the inclusion/exclusion criteria, and the data extracted and analyzed in this literature review paper. Section 3 elaborately discusses different thematic areas, and Section 4 presents the research gaps to address future research agendas. Section 5 presents a detailed discussion of the specified RQs. Lastly, Section 6 consists of the conclusion for this SLR.

## 2. Methodology

A systematic review is a research study that looks at many publications to answer a specific research topic. This study follows such a review to examine previous research studies that include identifying, analyzing, and interpreting all accessible information relevant to the recent progress of pertinent literature on Web 3.0 or Semantic Web in medical and healthcare or our phenomenon of interest. In the advancement of medical and healthcare analysis, numerous SLRs have been undertaken with inductive methodologies to identify major themes where Semantic Web technologies are being adopted [28, 29]. In our study, we adopted the procedures outlined by Keele with a few important distinctions to assure the study's transferability, dependability, and transparency, emphasizing and documenting the selection method [30]. The guidelines outlined in that paper were derived from three existing approaches used by medical researchers, two books written by social science researchers, and a discussion with other academics interested in evidence-based practice [8, 31–40]. The guidelines have been modified to include medical policies in order to address the unique challenges of software engineering research.

Our study sequentially conducted an SLR to accomplish the precise objectives. At first, we planned the necessary approach to identify the problems. Next, we collected related study materials and retrieved data from them. Finally, we documented the findings and carried out the research in the following steps (see Figure 1) maintaining its replicability as well as precision.

(i) ***Step 1***. Plan the review by finding appropriate research measures to detect corresponding documents.
(ii) ***Step 2***. Collect analyses by outlining the inclusion and exclusion criteria to assess their applicability.
(iii) ***Step 3***. Extract relevant data using numerous screening approaches to use accordingly.
(iv) ***Step 4***. Document the research findings.

*2.1. Planning the Review.* The very first stage in conducting SLR is to identify the needs for a specific systematic review, outline the research questions, design a procedural review, and offer a study framework to assist the investigation in subsequent phases to identify the systematic review's significant objectives. This phase begins with the identification of needs for the proposed systematic review. Section 1 of this paper went into detail about why a systematic review of Semantic Web technologies in healthcare was deemed necessary. Following that, the definition of research questions, the selection of a synthesis method, initial keywords, and databases are given. To begin, we devised the RQs for this SLR in order to gain a comprehensive understanding of the semantic-based solutions in the field of healthcare. Defining research questions is an important part of conducting a systematic review because they guide the overall review methodology. Based on the objectives, we conducted a pilot study of a systematic review of fifteen sample studies, resulting in the broad application of the Semantic Web to a specific niche, refinement of research questions, and redefinition of the review research protocol. To find relevant scientific contributions for our RQs, we used Scopus, IEEE Xplore Digital Library, ACM Digital Library, and Semantic Scholar. Furthermore, we utilized the primary term "Web 3.0 or Semantic Web" to search the databases and then identified and refined the comprehensive keywords that would be used as search strings. We did not limit our search to a single period instead; we looked at all linked studies.

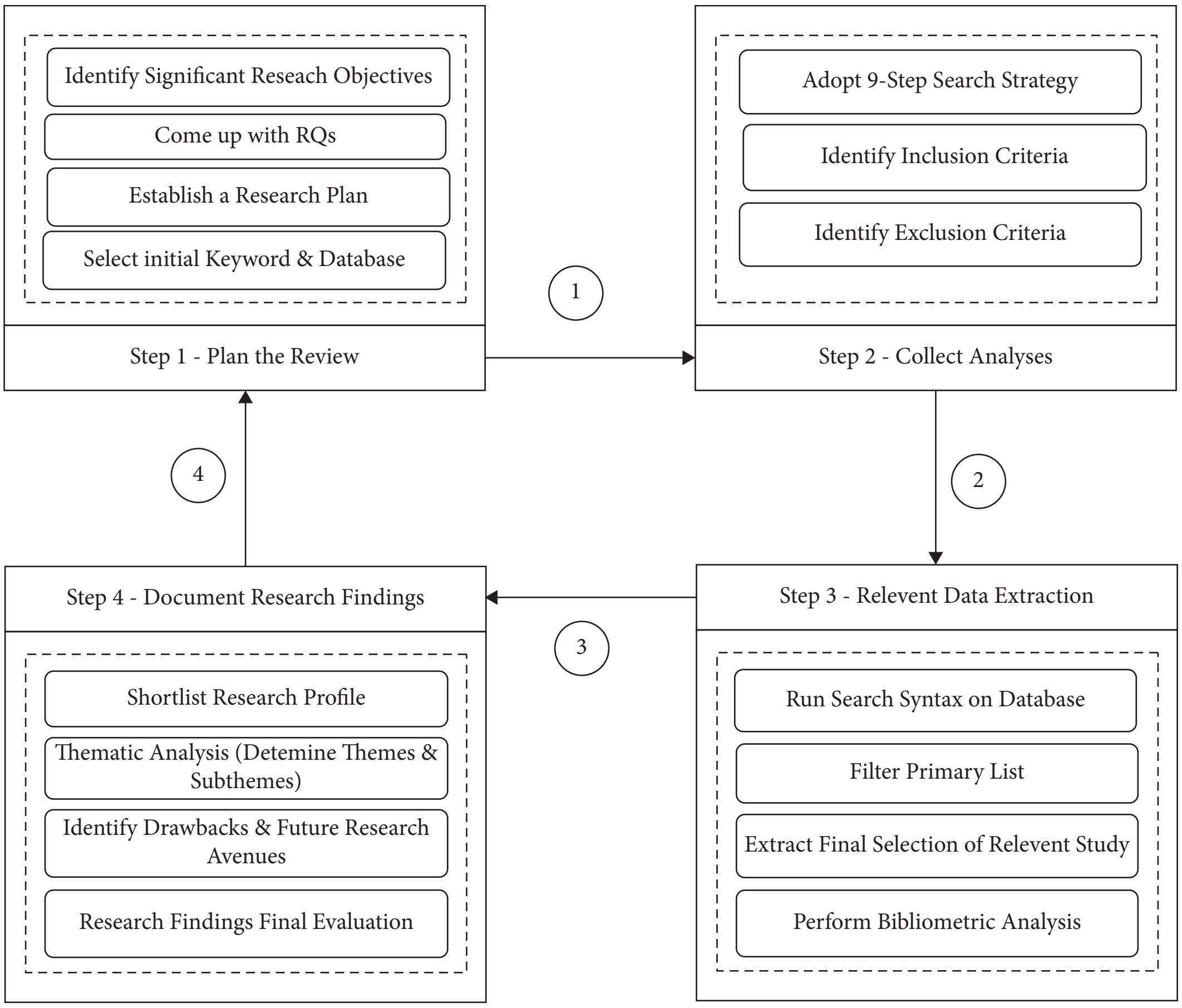

Figure 1: SLR methodology and protocols.

*2.2. Collecting Analyses.* A systematic review's unit of analysis is crucial since it broadens the scope of the overall approach. This study aims to better understand how Web 3.0 or Semantic Web technologies are employed in medical and healthcare settings, as well as to identify the extent to which they have been applied. We have selected academic research articles and journals as the unit of analysis for our SLR. We specified inclusion and exclusion criteria to narrow the investigation in the following study selection process, as shown in Table 1. To gather our search phrases, we used a nine-step procedure as mentioned in [41]. The studies obtained from online repositories were compared with exclusion criteria to select peer-reviewed papers and eliminate any non-peer-reviewed studies. To perform this review, we employed decisive exclusion criteria to identify grey literature, which included white papers, theses, project reports, and working papers. To remove language barriers, we only selected papers written in English. We did not consider any review papers or project reports to maintain the quality. Older publications that have never been cited were excluded from the review to explore the potential value of Web 3.0 and SW technologies in medical and healthcare.

*2.3. Extracting Relevant Data.* Initially, we searched for papers in Google Scholar with "Web 3.0 in medical and healthcare" keywords. However, reviewing the title and abstract from the top 50 articles further improved the search keyword to develop a more appropriate search string. The top search string ("Semantic Web" OR "Web 3.0") AND ("Healthcare" OR "medical") was used in Scopus, IEEE Xplore Digital Library, ACM Digital Library, and Semantic Scholar to find related papers for our SLR on 22 January 2022. We found a total of 4137 papers, including 2237 from IEEE Xplore Digital Library, 1761 from Scopus, 103 from Semantic Scholar, and 36 from ACM Digital library. Primary review grasped articles up to 2001. So, all the identified publications were from 2001 to 2021. Four authors performed the screening method through different stages. After each step, a discussion session was held to finalize the step and move further.

Table 1: Criteria for inclusion and exclusion.

| Inclusion criteria (IC) | Exclusion criteria (EC) |
|---|---|
| (**IC1**) Primary studies<br>(**IC2**) Peer-reviewed publications<br>(**IC3**) The studies are written in English language<br>(**IC4**) The research must be based on empirical evidence (qualitative and quantitative research)<br>(**IC5**) Journal articles published through January 22, 2022<br>(**IC6**) Studies available in full text<br>(**IC7**) Studies that focus on the Semantic Web to support medical and healthcare<br>(**IC8**) Any published study that has the potential to address at least one research question | (**EC1**) Studies not written in English<br>(**EC2**) White papers, working papers, positional papers, review papers, short papers (<4 pages), and project reports.<br>(**EC3**) Theses, editorials, keynotes, forum conversations, posters, editorials, analysis, tutorial overviews, technological articles, and essays.<br>(**EC4**) Grey literature, i.e., editorial, abstract, keynote, and studies without bibliographic information, e.g., publication date/type, volume, and issue number<br>(**EC5**) Research does not focus on the SW to support medical and healthcare |

At first, we checked for any duplicate articles from both indexing databases. We eliminated available duplicate articles by checking the Digital Object Identifier (DOI) and the research heading. After removing the duplicate articles, we were left with 1923 articles. After that, titles, keywords, and abstracts were read as part of the preliminary screening process. During the screening procedure, articles were divided into three categories: retain, exclude, and suspect. After removing articles unrelated to Web 3.0 or Semantic Web in medical and healthcare, only 1741 articles were retrained. Upon analyzing the contents of both suspect and retain studies using the inclusion and exclusion criteria listed in Table 1, we were left with 343 publications. Following that, we read the full text of the articles that were picked, and we were left with 54 papers being considered for our conclusive stage. Finally, we applied the snowballing strategy, also known as the citation chaining technique [42]. Surprisingly, this step resulted in the addition of another ten studies (7 from backward citation and three from forwarding citation). The final review pool thus comprised 65 papers being considered for our conclusive stage (Figure 2 depicts the study selection process in detail).

*2.4. Document Research Findings.* The shortlisted research papers were profiled using descriptive statistics, which include publication year, methodology, and publication sources [23, 43, 44]. According to the chronology of the number of publications, the majority of the research articles were published in 2013. However, between 2018 and 2021, the number declined. Figure 3 depicts the yearly (between 2001 and 2022) distribution of published papers.

The majority of the studies presented a framework for developing a medical data information management system. Web 3.0 technologies appear to be in their early phases of adoption, with scholars only recently becoming interested in the topic. A few other papers discussed medical data interchange mechanisms, diseases, frontier technology such as AI and NLP, and regulatory conditions. Nearly half of the research ($n = 39$) was published between 2001 and 2012, with the remaining studies ($n = 26$) published after that (see Figure 3). The Semantic Web theory gained widespread interest after the architect of the World Wide Web, Tim Berners-Lee, James Hendler, and Ora Lassila popularized it in a Scientific American article in May 2001 [15]. This trend also gained momentum in recent years, with John Markoff coining the term Web 3.0 in 2006 and Gavin Wood, Ethereum's co-founder, coining the word later in 2014.

Medical and healthcare writings have been published in several renowned conferences, journals, book series, and events. The 65 shortlisted papers are distributed throughout 27 conference proceedings, 21 journals, and 17 book series. The descriptive analysis depicts that 65 shortlisted analyses were authored by 25 publishers, accompanied by Springer ($n = 17$), IEEE Xplore ($n = 15$), IOS Press ($n = 6$), ACM ($n = 5$), and Elsevier ($n = 3$). Only a few publishers published many studies. The reset included 15 publishers, each of whom only published one study. However, the majority of the papers were published in Lecture Notes in Computer Science (LNCS), CEUR Workshop Proceedings, and Studies in Health Technology and Informatics Series (see Figure 4). Furthermore, our SLR demonstrates the wide geographic span of existing research papers. The United States (11 articles), France (23 articles), India (9 articles), Canada (8 articles), Belgium (4 articles), and South Korea (4 articles) all had a significant number of studies. Figure 5 summarizes the past literature's geographical distribution.

According to the systematic literature review, the application of Semantic Web technologies in the field of healthcare is a prominent classical research theme, with many innovative and promising research topics. The number of Semantic Web publications and interest in healthcare has increased rapidly in recent years, and

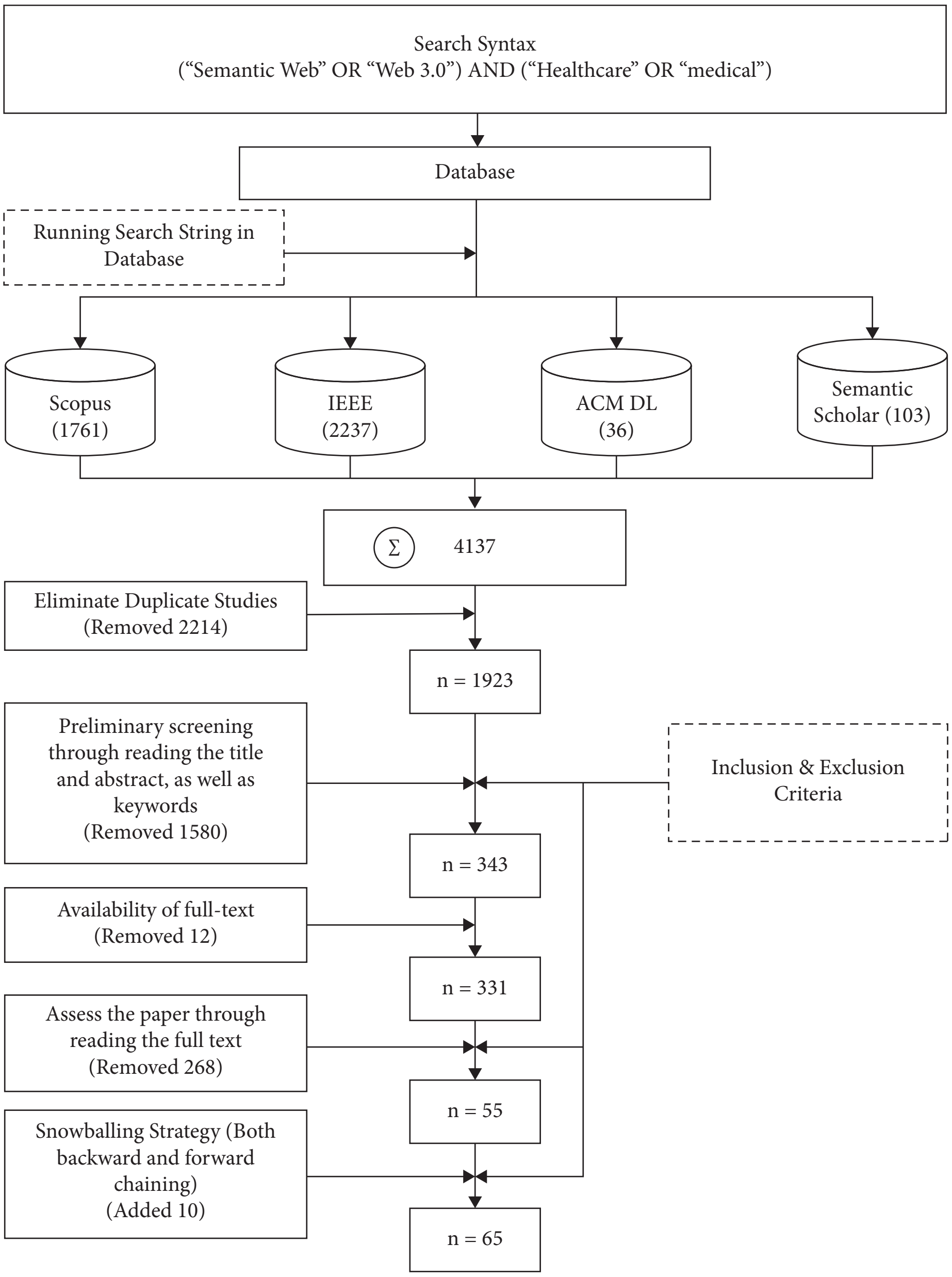

Figure 2: Study selection process.

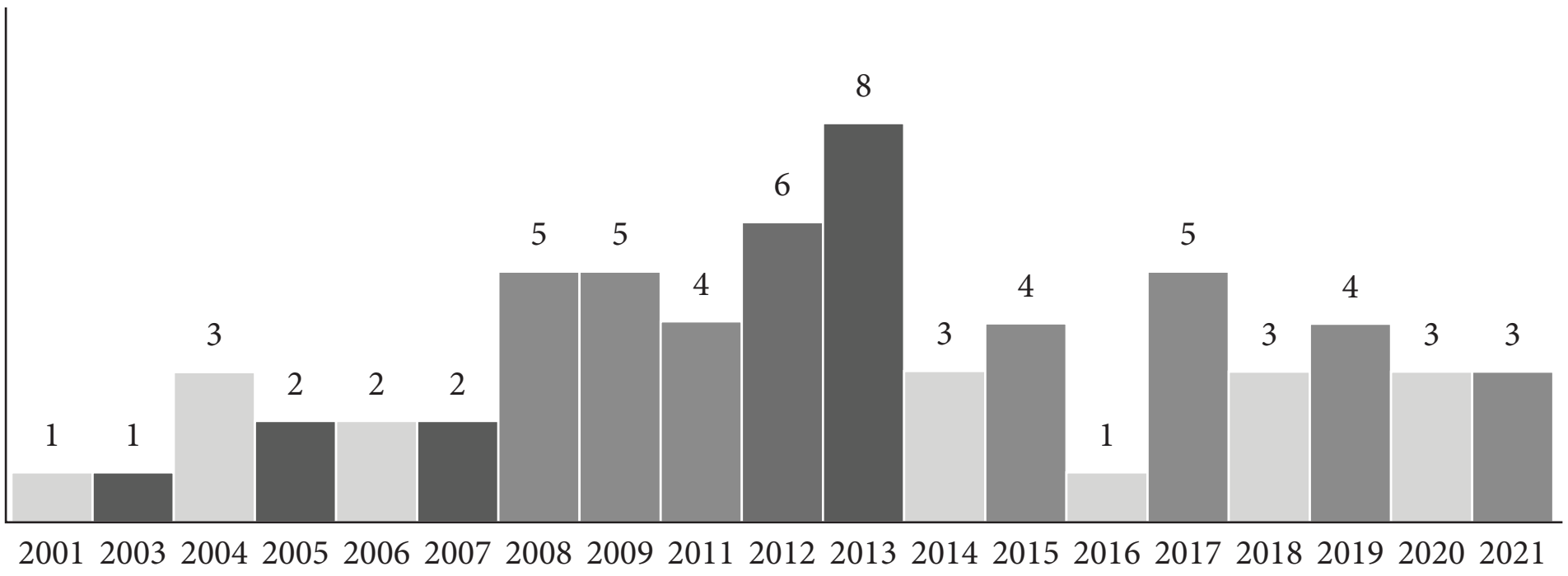

Figure 3: Number of articles published yearly.

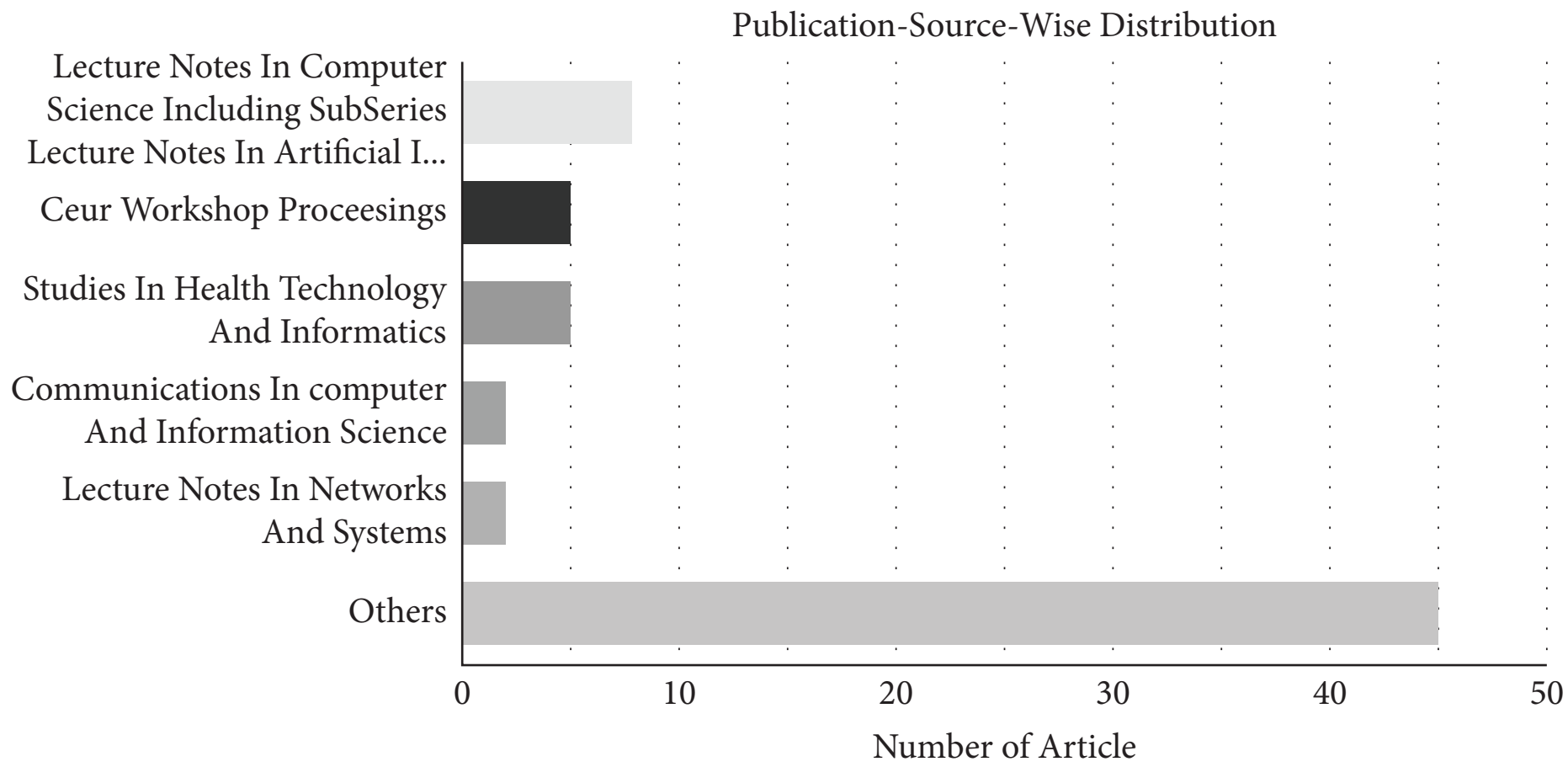

Figure 4: Publication-source-wise distribution.

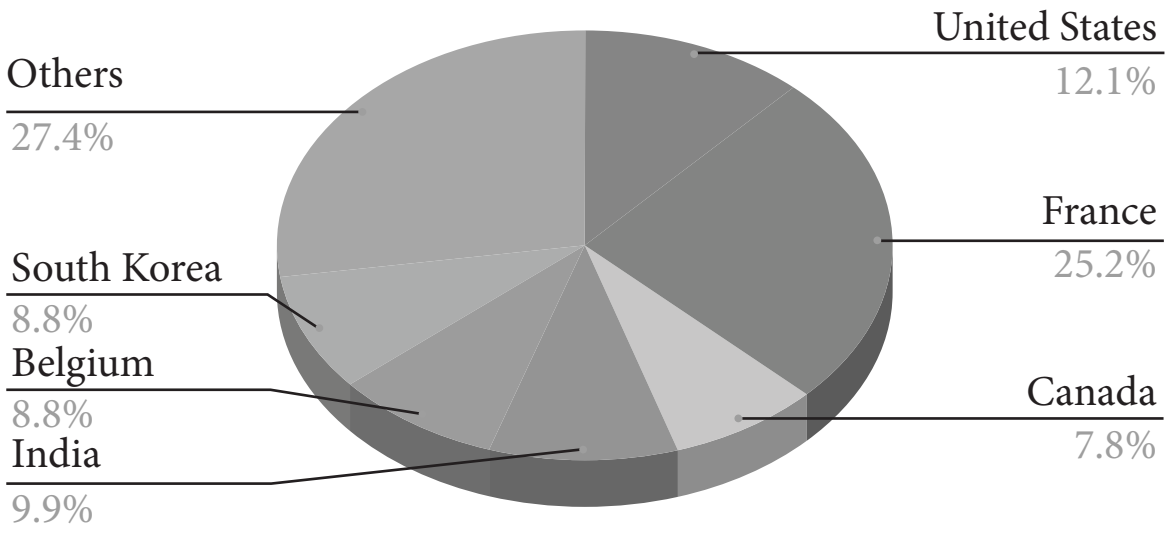

Figure 5: Country-wise article distribution.

Semantic Web methods, tools, and languages are being used to solve the complex problems that today's healthcare industries face. Semantic Web technology allows comprehensive knowledge management and sharing, as well as semantic interoperability across application, enterprise, and community boundaries. This makes the Semantic Web a viable option for improving healthcare services by improving tasks such as standards and interoperable rich semantic metadata for complex systems, representing patient records, investigating the integration of Internet of Things and artificial computational methods in disease identification, and outlining SW-based security. While there are interesting possibilities for the application of Semantic Web technologies in the healthcare setting, some limitations may explain why those possibilities are less apparent. We believe one reason is a lack of support for developers and researchers. Semantic Web-based healthcare applications should be viewed as independent research prototypes that must be implemented in real-world scenarios rather than as a widget that is integrated with the Web 2.0-based solution. This study discusses the findings and future directions from two different perspectives. First, consider the potential applications of Semantic Web technologies in different healthcare scenarios and also look at the barriers to their practical application and how to overcome them (see Section 3). Last, the fourth (see Section 4) section discusses the scope of research in Semantic Web-enabled healthcare.

## 3. Analysis of the Selected Articles: Thematic Areas

This section focuses on three key steps: summarizing, comparing, and discussing the shortlisted papers to describe and categorize them into common themes. To systematically analyze all 65 studies, we adopted the technique used in recently published SLRs [23, 43]. After identifying and selecting relevant papers that could answer our research questions, we used the content analysis technique to classify, code, and synthesize the findings of those studies. A three-step approach was proposed by Erika Hayes et al., which was used to interpret unambiguous and unbiased meaning from the content of text data [45]. The steps were as follows: (a) the authors assigned categories to each study and a coding scheme created directly and inductively from raw data using valid reasoning and interpretation; (b) the authors immersed themselves in the material and allowed themes to arise from the data to validate or extend categories and coding schemes using directed content analysis; (c) the authors used summative content analysis, which begins with manifesting content and then expands to identify hidden meanings and themes in the research areas.

This thematic analysis answers the second research question (RQ2), "What are the primary objectives of using the Semantic Web, and what are the major areas of medical and healthcare where Semantic Web technologies are adopted?", and this analysis architecture highlights five broad medical and healthcare-related research themes based on their primary contribution (see Table 2), notably e-healthcare service, diseases, information management, frontier technology, and regulatory conditions.

Two themes, namely, IoT and cloud computing, were nevertheless left out since they lack a wide description that would be useful in developing a meaningful theme. Some of the papers from which we defined these two thematic areas were included in the selected themes based on their similarity to the chosen thematic areas. Figure 6 illustrates this categorization, with different themes' description, which emerged from our review.

Table 2: Derived themes and their descriptions.

| *Theme name* | Theme description |
|---|---|
| *E-healthcare service* | E-healthcare services are defined as healthcare services and resources that are improved or supplied over the Internet and other associated technologies to reduce the burden on the patients. |
| *Diseases* | Diseases include a wide range of illnesses, including dementia, diabetes, chronic disorders, cardiovascular disease, and critical limb ischemia. The objective is to use SWT to integrate medical information and data from various electronic health data sources for efficient diagnosis and clinical services. |
| *Information management* | Information management in healthcare is the process of gathering, evaluating, and preserving medical data required for providing high-quality healthcare management systems. In this thematic area, we discuss how SWT can be used to develop the management of massive healthcare data. |
| *Frontier technology* | In a broad sense, frontier technology in healthcare refers to technologies such as artificial intelligence, various spectrum of IoT, augmented reality, and genomics that are pushing the boundaries of technological capabilities and adoption. Only the works of scholars who collaborated with Semantic Web and frontier technologies to meet healthcare demand are included in this category. |
| *Regulatory conditions* | Regulatory conditions refer to the activities that aim to develop adequate underlying motives and beliefs, guidelines, and healthcare protocols across healthcare facilities and systems. This section's research focuses on the improvement of good practice and clinical norms using SWT for documenting the semantics of medical and healthcare data and resources. |

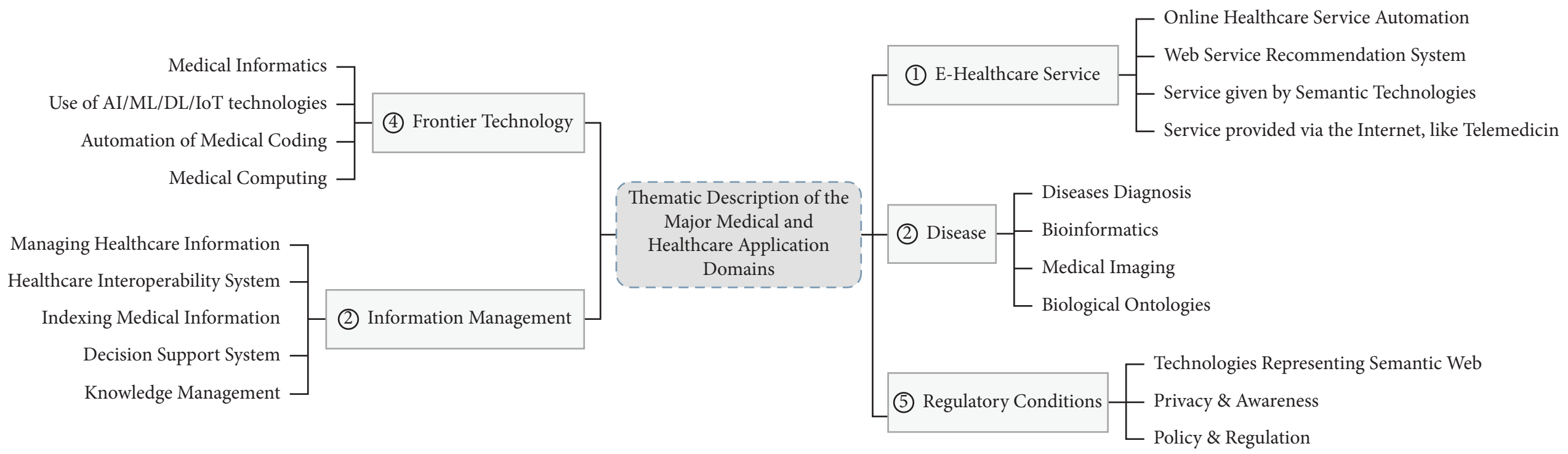

Figure 6: Thematic description of Semantic Web approaches in medical and healthcare.

*3.1. E-Healthcare Service.* The use of various technologies to provide healthcare support is known as e-service in healthcare or e-healthcare service. While staying at home, a person can obtain all the necessary medical information as well as a variety of healthcare services such as disease reasoning, medication, and recommendation through e-healthcare services. It is similar to a door-to-door service. The Semantic Web or Web 3.0 plays a critical role in this regard. The Semantic Web offers a variety of technologies, including semantic interoperability, semantic reasoning, and morphological variation that can be used to create a variety of frameworks that improve e-healthcare services.

SW makes the task of sharing medical information among healthcare experts more efficient and easier [2, 46–48]. A dataset information resource for medical knowledge makes the work more trouble-free and faster. A healthcare dataset information resource has been created along with a question-answering module related to health information [26]. Combining different databases can be more effective as it expands the information range of knowledge. In this respect, Barisevičius et al. [49] designed a medical linked data graph that combines different medical databases and they also developed a chatbot using NLP-based knowledge extraction that provides healthcare services by supplying knowledge about various medical information. Besides information sharing and database combining, Semantic Web-based frameworks can provide virtual medical and hospital-based services. A system has been created that provides medical health planning according to patient's information [50]. Concerning this, it could be very helpful if there is a system that can match patient requirements with the services. Such a matchmaking system has been developed to match the web services with the patient's requirements for medical appointments [51]. To provide hospital-based services, a Semantic Web-based dynamic healthcare system was developed using ontologies [17]. Disease reasoning is a vital task for e-healthcare services. A number of frameworks were developed that are used for reasoning diseases [49, 52, 53]. In addition, some authors implemented systems that provide support for sequential decision making [54–57]. Moreover, Mohammed and Benlamri [21] designed a system that could help to prescribe differential diagnosis recommendations. Grouping similar diagnosis patients can be useful to enhance the medication process. In this regard, Fernández-Breis et al. [58] created a framework to group the patients by identifying patient cohorts. Moreover, Kiourtis et al. [59] proposed a new device-to-device (D2D) protocol for short-distance health data exchange between a healthcare professional and a citizen utilizing a sequence of Bluetooth communications. Supplying medical information to people is one of the main tasks of e-healthcare services [58]. Before proceeding with a medical diagnosis, we need to be sure about the correctness of the procedure. Andreasik et al. [60] developed a Semantic Web-based framework to determine the correctness of medical procedures. Various systems for medical education were developed using Semantic Web technologies such as a web service delivery system [4], a web service searching system [61], and an e-learning framework for the patients to learn about different medical information [62, 63]. Some articles discussed the rule-based approaches for the advancement of medical applications [64, 65]. Quality assurance of Semantic Web services is necessary, and so a framework was created using a Semantic Web-based replacement policy to assure the quality of a set of services and replace it with a newly defined subset of services when the existing one fails in execution [3]. A framework was designed for Semantic Web-based data representation [66]. Meilender et al. [67] described the migration of Web 2.0 data into Semantic Web data for the ease of further advancement in Web 3.0.

Researchers used different Semantic Web services to convert the relational database to create Resource Description Framework (RDF) and Web Ontology Language (OWL)-based ontologies. It is done by extracting the instances from the relational databases and representing them into RDF datasets [21, 55, 57, 62]. In some prior literature, many RDF datasets were created using Apache JENA 4.0 [4], different versions of protégé were used to construct and represent various healthcare ontologies [2, 17], Apache Jena framework was used for OWL reasoning on the RDF datasets [50, 53], and the EYE engine was used for reasoning [54]. Besides, Kiourtis et al. [68] developed a technique for converting healthcare data into its equivalent HL7 FHIR structure, which principally corresponds to the most used data structures for describing healthcare information. Furthermore, a sublanguage of F-logic named Frame Logic for Semantic Web Services (FLOG4SWS) and web services along with some features of Flora-2 was used to represent the ontology [51]. The authors of some papers used RDF and OWL for data representation of different ontologies [50, 52, 54, 66]. Mohammed and Benlamri [21] offered a number of Semantic Web strategies for ontology alignment, such as ontology matching and ontology linking, and some used ontology mapping for the ontology alignment [58, 66]. By combining RDF and semantic mapping features, Perumal et al. [69] provided a translation mechanism for healthcare event data along with Semantic Web Services and decision making. In addition, a linked data graph (LDG) is utilized to combine numerous publicly available medical data sources using RDF converters [49]. The works in [52, 54] used Notation3 for data mapping. SPARQL was used as the query language for the database [2, 17, 50, 52, 57]. Besides, the Jena API was also used as a query language [21]. The Semantic Web's rules and logic were expressed in terms of OWL concepts using the Semantic Web Rule Language (SWRL) [55, 57]. TopBraid Composer is used as the Semantic Web modeling tool [60].

There was no proof that the system created using semantic networks was able to share knowledge among healthcare services [2, 46, 48]. Researchers did not mention how a system can be integrated with different types of datasets in the world [2, 47]. In their paper, Ramasamy et al. [3] did not mention whether the system could replace all types of services or not. Shi et al. [26] did not discuss the success rate of the datasets in their dataset information resource and the accuracies of different systems created with these datasets. No proper evaluation techniques have been given for linked data graph [49], Semantic Web service delivery systems [4, 50], and Semantic Web reasoning system [52, 53] in their studies. There is no discussion of the reliability and validity of numerous decision making and recommendation systems [21, 54, 70]. Podgorelec and Gradišnik [64] did not provide information about the betterment of the combined Semantic Web technologies and rule-based systems against other alternatives. Most of the articles discussed or offered various techniques to build different healthcare services, but there are only a few articles that implemented the proposed systems and tested them in a real-life context.

*3.2. Diseases.* This thematic area aims to specifically identify and discuss the contributions of Semantic Web technologies to reach interoperability of information in the healthcare sector and aid in the initial detection and nursing of diseases, such as diabetes, chronic conditions, cardiovascular disease, dementia, and so on. SW provides a framework to integrate medical knowledge and data for effective diagnosis and clinical service. They help to select patients, recognize drug effects, and analyze results by using electronic health data from numerous sources. The queryMed packages were proposed for pharmaco-epidemiologists that link medical and pharmacological knowledge with electronic health records [10]. This application searches for people with critical limb ischemia (CLI) with at least one medication or none at all and gives them healthcare recommendations. SW also emphasizes the study of phenotypes and their influence on personal genomics. The Mayo Clinic's project, Linked Clinical Data (LCD), facilitates the use of SW and makes it easier to extract and express phenotypes from electronic medical records [71]. It also emphasizes the use of semantic reasoning for the identification of cardiovascular diseases. Besides this, it aims to improve healthcare service quality for people suffering from chronic conditions. Proper planning and management are required for the better treatment and management of chronic diseases. Thus, the Chronic Care Model (CCM) provides knowledge-based acquisition to patients [72].

Ontology-based applications such as the Concept Unique Identifier (CUI) from Unified Medical Language System, Drug Indication Database (DID), and Drug Interaction Knowledge Base (DIKB) are widely used in the medical domain to establish mappings between medical terms [10]. In the context of ontology, the ECOIN framework uses a single ontology, multiple view approach that exploits modifiers and conversion functions for context mediation between different data sources [73]. To support clinical knowledge sharing through interaction models, the OpenKnowledge project has been initiated from different data sources [9], and K-MORPH architecture has been proposed for a unified prostate cancer clinical pathway.

Along with information sharing, medical data management is critical in the diagnosis of disorders like dementia. To establish a better diagnosis method for dementia, a medical information management system (MIMS) was designed using SW technologies through the extraction of metadata from medical databases [74, 75]. In order to further eliminate the e-health information and knowledge sharing crisis, Bai and Zhang [76] suggested Integrated Mobile Information System (IMIS) for healthcare. It provides a platform to connect diabetic patients with care providers to receive proper treatment and diagnosis facilities at home. The Diabetes Healthcare Knowledge Management project also aims to ease decision support and clinical data management in diabetes healthcare processes [72].

To construct decision models for the Diabetes Healthcare Knowledge Management framework, tools such as Semantic Web Rule Language (SWRL), OWL, and RDF were used. This ontology-based knowledge framework provides ontologies, patient registries, and an evidence-based healthcare resource repository for chronic care services [72]. Web Ontology Language (OWL), Resource Description Framework (RDF), and SPARQL were also commonly used for the creation of metadata in dementia diagnosis [77]. On the other hand, the Semantic Web-based retrieval system for the pathology project, known as "A Semantic Web for Pathology," involves building and managing ontology for lungs which was made up of common semantic tools RDF and OWL which were used along with RDQL query language [20].

Even though effective frameworks were proposed to diagnose certain diseases, research gaps still exist that affect medical data management. For instance, the fuzzy technique-based service-oriented architecture has proved to be beneficial in terms of adjustability and reliability. But still, in the context of domain-specific ontologies, the applicability of this architecture is yet to be validated [78]. Effective distribution of knowledge into the existing healthcare system is a huge challenge in augmenting decision making and improving the care service quality. Therefore, future works are intended to focus on embedding knowledge and conducting user evaluations for better disease management.

*3.3. Information Management.* Managing patients' information and storing test results are significant tasks in the medical and healthcare industries. The application of the SW-based approach in this area can make an influential impact on this data organization. Such an approach to gather valuable and new medical information was primarily made by creating a network of computers [79]. Domain ontology was created according to the user's choice, suggesting medical terminologies to retrieve customized medical information [80]. RDF datasets can be used to find the trustworthiness of intensive care unit (ICU) medical data [70]. The SW has also been used to document healthcare video contents [81] and radiological images to provide appropriate information about those records [82].

However, moving from the conventional web-based information management to the Semantic Web had some reasons. As medical knowledge is essential to verify and share across hospitals and medical centers, introducing the Semantic Web approach helped to achieve a proper mapping system [83]. A medical discussion forum based on the SW helped to exchange valuable data among healthcare practitioners to map-related information in the dataset [84]. The use of the fuzzy cognitive system in the SW also helped to share and reuse knowledge from databases and simplify maintenance [85]. This methodology also helped to improve data integration, analysis, and sharing between clinical and information systems and researchers [86]. Moving towards this approach also aided the researchers in connecting different data storage domains and creating effective mapping graphs [87].

Though the approach of SW in healthcare has a broad area, most applications are pretty similar. The framework mainly proposed the use of RDF, SPARQL, and OWL [4, 76]. Link relevance methods were used to produce semantically relevant results to extract pertinent information from domain knowledge [49]. Ontology-based logical framework procedures and SMS architecture helped to organize the heterogeneous domain network [88, 89].

Evaluating the system's performance is necessary to get the actual results. A Health Level 7 (HL7) messaging mechanism has been developed for mapping the generated Web Service Modeling Ontology [90]. However, there were some issues regarding the heterogeneity problem. JavaSIG API was used to generate the HL7 message to resolve these issues [91]. Some of the evaluation tools are not advanced enough to handle vast amounts of data. PMCEPT physics algorithms were used to verify the algorithm [92]. Abidi and Hussain [9] created two levels to characterize different ontological models to establish morphing. BioMedLib Search Engine creation for economic efficiency helped to develop a Semantic Web framework for rural people [25]. The Metamorphosis installation wizard converted the text format UMLS into a MySQL database UMLS in order to access a SPARQL endpoint [93].

However, the frameworks proposed in different statements were not implemented precisely, which created a gap in each framework. Some frameworks are proposed to integrate with the blockchain for additional security and privacy [23, 94–96]. AI and IoT integration can also enhance system maintenance [1]. Hussain et al. [97] suggested a framework named Electronic Health Record for Clinical Research (EHR4CR), but they did not get any actual results from this framework in the real world [97]. The proposed framework's implementation result will provide more development on this.

*3.4. Frontier Technology.* In this segment, we critically analyze works that are primarily keen on how cutting-edge technologies like AI and computer vision can be applied to the medical field with the continuous advancement of science and technology. Semantic Web-enabled intelligent systems leverage a knowledge base and a reasoning engine to solve problems, and they can help healthcare professionals with diagnosis and therapy. They can assist with medical training in a resource-constrained environment. To illustrate, Haque et al. [8], Chondrogiannis et al. [98], Haque and Bhushan [99], and Haque et al. [24] created a secure, fast, and decentralized application that uses blockchain technologies to allow users and health insurance organizations to reach an agreement during the implementation of the healthcare insurance policies in each contract. To preserve the formal expression of both insured users' data and contract terms, health standards and Semantic Web technologies were used. Accordingly, significant work has been proposed by Tamilarasi and Shanmugam [100] which explores the relationship between the Semantic Web, machine learning, deep learning, and computer vision in the context of medical informatics and introduces a few areas of applications of machine learning and deep learning algorithms. This study also presents a hypothesis on how image as ontology can be used in medical informatics and how ontology-based deep learning models can help in the advancement of computer vision.

The real-world healthcare datasets are prone to missing, inconsistent, and noisy data due to their heterogeneous nature. Machine learning and data mining algorithms would fail to identify patterns effectively in this noisy data, resulting in low accuracy. To get these high-quality data, data preprocessing is essential. Besides, RDF datasets representing healthcare knowledge graphs are very important in data mining and integrating IoT data with machine learning applications [8, 101]. RDF datasets are made up of a distinguishable RDF graph and zero or more named graphs, which are pairings of an IRI or blank node with an RDF graph. While RDF graphs have formal model-theoretic semantics that indicate which world configurations make an RDF graph true, there are no formal semantics for RDF datasets. Unlike traditional tabular format datasets, RDF datasets require a declarative SPARQL query language to match graph patterns to RDF triples, which makes data preprocessing more crucial. In the context of data preprocessing, Monika and Raju [101] proposed a cluster-based missing value imputation (CMVI) preprocessing strategy for preparing raw data to enhance the imputed data quality of a diabetes ontology graph. The data quality evaluation metrics R2, D2, and root mean square error (RMSE) were used to assess simulated missing values.

Nowadays, question-answering (QA) systems (e.g., chatbots and forums) are becoming increasingly popular in providing digital healthcare. In order to retrieve the required information, such systems require in-depth analysis of both user queries and records. NLP is an underlying technology, which converts unstructured text into standardized data to increase the accuracy and reliability of electronic health records. A Semantic Web application has been deployed for question-answering using NLP where users can ask questions about health-related information [27]. In addition, this study introduces a novel query simplification methodology for question-answering systems, which overcomes issues or limitations in existing NLP methodologies (e.g., implicit information and need for reasoning).

The majority of contributions to this category have organized their work using semantic languages on a smaller scale. Besides, it is noteworthy that hardly any of the approaches, except [27, 101], adopted a framework for developing their models. Asma Ben et al. used a benchmark (corpus for evidence-based medicine summarization) to evaluate the question-answering (QA) system and analyzed the obtained outcomes [27]. Some studies have not included a prior literature review for the discovery of available frontier services [100]. In addition, the study shows that with the soaring demand for better, speedier, more accurate, and personalized patient treatment, deep learning powered models in production are becoming increasingly prevalent. Often these models are not easily explainable and prone to biases. Explainable AI (XAI) has grown in popularity in healthcare due to its extraordinary success in explaining decision-making criteria to systems, reducing unintended outcomes and bias, and assisting in gaining patients' trust—even when making life-or-death decisions [102]. To the best of our knowledge, XIA has gleaned attention on ontology-based data management but received relatively little attention on collaborating Semantic Web technologies across healthcare, biomedical, clinical research, and genomic medicine. Similarly, within the IoT system spectrum, invocation of semantic knowledge and logic across various Medical Internet of Things (MIoT) applications, gathering vast amounts of data, monitoring vital body parameters, and gathering detailed information from sensors and other connected devices, as well as maintaining safety, data confidentiality, and service availability also received relatively little attention.

*3.5. Regulatory Conditions.* This segment concentrates on Semantic Web-based tools, technologies, and terminologies for documenting the semantics of medical and healthcare data and resources. As the healthcare industries generate a massive amount of heterogeneous data on a global scale, the use of a knowledge-based ontology on such data can reduce mortality rate and healthcare costs and also facilitate early detection of contagious diseases. Besides, the SW provides a single platform for sharing and reusing data across apps, companies, and communities. The biomedical community has specific requirements for the Semantic Web of the future. There are a variety of languages that can be used to formalize ontologies for medical healthcare, each with its expressiveness. A collaborative effort led by W3C, involving many research and industrial partners, set the requirements of medical ontologies. A real ontology of brain cortex anatomy has been used to assess the requirements stated by W3C in two available languages at that time, Protégé and DAML + OIL [103]. The development and comparative analysis contexts of brain cortex anatomy ontologies are partially addressed in this. In 2019, a survey-based study was conducted to determine faculty and researcher usage, impact, and satisfaction with Web 3.0 networking sites on medical academic performance [104]. This study explores the awareness and willingness to implement Web 3.0 technologies within healthcare at Rajiv Gandhi University of Health Sciences. The results of this study imply that Web 3.0 technologies have an impact on professor and researcher academic performance, with those who are tech-savvy being disproportionately found in high-income groups [104].

Documentation of semantic tools and data is required to resolve healthcare reimbursement challenges. Besides, regulations are also necessary to standardize semantic tools while ensuring that healthcare communities and systems adhere to general health policies. Unfortunately, we found only a few works focusing on this challenge based on SWT. Only the study conducted by Sugihartati [104] adopted a proper survey methodology. Therefore, future efforts should focus on regulating, documenting, and standardizing semantic tools, technologies, and health resources, as well as conducting user evaluations to understand and optimize functional efficiency and accelerate market access for medicines for general health.

Tables 3–5 provide a detailed analysis of the studied works for the derived five categories.

## 4. Research Gaps

This systematic literature review presents a vast knowledge about the use of Web 3.0 or Semantic Web technology in different approaches to the medical and healthcare sector. By analyzing various kinds of literature, we recognized different research gaps to address future research avenues, which will enable scholars from different parts to examine the area and discover new developments. Table 4 summarizes the overall research gaps and Table 5 summarizes the future research avenues we encounter during the literature review.

*4.1. Scope of E-Healthcare Service Research.* Even though studies in the domain of e-healthcare services suggested and created numerous frameworks to provide vital support to the users, there are still research gaps among the methods. Several frameworks were proposed to facilitate data interoperability. However, based on what we know best, none of the proposed frameworks has been implemented in the actual world. Furthermore, there is no evidence of knowledge sharing among organizations using semantic network-based systems. Besides, just a handful of the research papers included assessment methodologies and a discussion of the findings. Furthermore, the frameworks that provide medical services such as disease reasoning, decision making, and drug recommendations lack reliability and validity. Most of the research articles suggested architectures but did not implement them, and their intended prototypes were never built.

*4.2. Scope of Disease Research.* Semantic Web technologies are being used in the healthcare sector to improve information interoperability and aid in identifying and treating diseases. Only a few studies among the 65 papers have examined the various frameworks for developing a fully functional system for either diabetic healthcare or disease collection of prebuilt queries. Earlier research also lacks mapping triplets of one illness RDF to other existing medical services, applications, and administrations. Researchers also lack the creation of intelligent user interfaces that grasp the semantics of clinical data. This paper shows that more study is required to efficiently use ontology in the healthcare sector to preserve data with proper evaluation criteria.

*4.3. Scope of Information Management Research.* Medical data are considered valuable information utilized to assist patients in receiving better care. It is challenging to implement Semantic Web technologies to store and search for data. Various studies attempt to adopt specific methods that may aid in the proper management of medical information; however, some gaps remain. There is no attempt to index high-quality videos and collect attributes for categorizing them. A validation gap exists due to the lack of suitable evaluation techniques. In most studies, RDF ontologies are used to collect information from websites and represent those data. However, no information is provided about how effective those models are in real-world applications.

*4.4. Scope of Frontier Technology Research.* Even though cutting-edge technology such as AI, ML, robotics, and the IoT has revolutionized the healthcare industry and helped improve everything from routine tasks to data management and pharmaceutical development, the industry is still evolving and looking for ways to improve. If we consider the aspect of research, the history of the Semantic Web and frontier technology is technically not new at all, yet the Semantic Web presents some limitations. Since the web began as a web of documents, converting each document into data is incredibly challenging. Various tools and approaches, such as natural language processing (NLP), may be used to do this task, but it would take a long time. However, only a small attempt has been made to integrate NLP and domain knowledge induction. Ontology and AI, and logic, have always been and will continue to be essential elements of AI development. Besides, connecting ontology to AI is frequently a problem in and of itself. Furthermore, because ontology trees often have a large number of nodes, real-time execution is problematic. Earlier studies have apparently failed to solve this problem. There have been significant attempts to incorporate the various aspects of IoT resources into ontology creation, such as connectivity, virtualization, mobility, energy, or life cycle [108, 109]. The authors attempted to enhance the computerization of the health and medical industry by utilizing the Internet of Things (IoT) and Semantic Web technologies (SWTs), which are two key emerging technologies that play a significant role in overcoming the challenges of handling and presenting data searches in hospitals, clinics, and other medical establishments on a regular basis. Despite its significant efforts to collaborate different IoT spectrum and Semantic Web technologies, research gaps in medical data management persist. For instance, after its introduction, the Medical Internet of Things (MIoT) has taken an active role in improving the health, safety, and care of billions of people.

Table 3: Summarization of the research contribution of the selected articles.

| Themes | Contributions |
|---|---|
| ***E-healthcare service*** | (i) An ontology-based semantic server for healthcare organizations to exchange information among them [2].<br>(ii) Discussed healthcare data interoperability and integration plan of the solution [46].<br>(iii) Used Semantic Web terms (SWT) to provide oral medicine knowledge and information [47] and to build a decision support system [56].<br>(iv) Developed a prototype that generates the desired reports using a high degree of data integration and discussed a production rule-based approach to establish a link between prevalent diseases and the range of the diseases in a particular gene [64].<br>(v) Represented global ontology via bridge methods to avoid conflicts among different local ontologies [65].<br>(vi) Implemented a WSMO (Web Service Modeling Ontology) automated service delivery system [57].<br>(vii) Designed a system for automatic alignment of user-defined EHR (electronic health record) workflows [4].<br>(viii) Proposed an upper-level-ontological service providing a mechanism to provide integrity constraints of data and to improve the usability of the medical linked data graph (LDG) services [49].<br>(ix) Developed a chatbot and a triaging system that provides information about diseases, screens users' problems, and sorts patients into groups based on the user's needs [49].<br>(x) Developed a healthcare dataset information resource (DIR) to hold dataset information and respond to parameterized questions [26].<br>(xi) A healthcare service framework that coordinates web services to locate the closest hospital, ambulance service, pharmacy, and laboratory during an emergency [17].<br>(xii) Used web service replacement policy to build a Semantic Web service composition model which replaces a set of services with a generated service subset when the previous set of services fails in execution [3].<br>(xiii) Proposed ontology-based data linking to understand and extract medical information more precisely [2].<br>(xiv) Integrated knowledge with clinical practice to provide guidelines in medicine [21].<br>(xv) An abstraction method that converts XML-type medical information to RDF and OWL to create electronic health record (EHR) architecture for the identification of patient cohorts [58].<br>(xvi) Designed a platform for solving complex medical tasks by interpreting algorithms and meta-components [66].<br>(xvii) Provided a strategy for suggestions in view of clients' likeness figuring and exhibited the adequacy of the model suggested through configuration, execution, and examination in social learning environments [61].<br>(xviii) Constructed semantic relationships of input and output medical-related parameters to resolve conflicts and algorithms that remove the redundancy of web service paths [54].<br>(xix) Used a management time and run time subsystem to discover the potential web services [62].<br>(xx) Integrated weak inferring with a single and explanation-based generalization to leverage the complementary strengths [53]. |
| ***Diseases*** | (i) Created an ontology to build and manage information about a particular disease [74].<br>(ii) Developed a web-based prototype of Integrated Mobile Information System for healthcare of diabetic patients [76].<br>(iii) Implemented embedded feedback between users and designers and communication mechanisms between patients and care providers [20].<br>(iv) QueryMed *R* package made the integration of clinical and pharmacological information that is used to distinguish all the medications endorsed for critical limb ischemia (CLI) and to recognize one contraindicated solution for one patient [10].<br>(v) A semantics-driven system based on EMRs that can break down multifactorial phenotypes, like peripheral arterial disease and coronary heart disease [71].<br>(vi) Discussed a way to deal with a unified prostate cancer clinical pathway by incorporating three different clinical pathways: Halifax pathway, Calgary pathway, and Winnipeg pathway [72].<br>(vii) Demonstrated the achievability and tolerability of a distributed web-oriented environment as an effective study and approval technique for characterizing a real-life setting [78]. |

Table 3: Continued.

| Themes | Contributions |
|---|---|
| ***Information management*** | (i) Proposed a brief process of integration for interoperability and scalability to create an ontology of inflammation [89].<br>(ii) Discussed an indexing mechanism to extract attributes from an audio-visual web system [81].<br>(iii) Developed ontology-enabled security enforcement for hospital data security [79].<br>(iv) Semantic Web mining-based ontologies allow medical practitioners to have better access to the databases of the latest diseases and other information [79].<br>(v) Proposed a medical knowledge morphing system to focus on ontology-based knowledge articulation and morphing of diverse information through logic-based reasoning with ontology mediation [105].<br>(vi) The annotation image (AIM) ontology was created to give essential semantic information within photos, allowing radiological images to be mined for image patterns that predict the biological properties of the structures they include [82].<br>(vii) Described a semantic data architecture where an accumulative approach was used to append data sources [70].<br>(viii) Implemented a functional web-based remote MC system and PMCEPT code system, as well as a description of how to use a beam phase space dataset for dosimetric and radiation therapy planning [92].<br>(ix) Discussed an approach using Notation3 (N3) over RDF to present a generic approach to formalizing medical knowledge [85].<br>(x) It was demonstrated that in the healthcare domain, knowledge management approaches and the synergy of social media networks may be used as a foundation for the creation of information system (IS). This helps to optimize data flow in healthcare processes and provides synchronized knowledge for better healthcare decision making (cardiac diseases) [106].<br>(xi) Using semantic mining principles, the authors described a technique for minimizing information asymmetry in the healthcare insurance sector to assist clients in understanding healthcare insurance plans and terms and conditions [22].<br>(xii) Discussed a mapping-based approach to generate Web Service Modeling Ontology (WSMO) description from HL7 (Health Level 7) V3 specification where Messaging Modeling Ontology (MMO) is mapped with WSMO [90].<br>(xiii) Designed a web crawler-based search engine to gather medical information as per patients' needs [80].<br>(xiv) A framework where patients can get relevant medical information from a personalized database, where the patient's medical history and current health condition are captured and then analyzed to search for particular information regarding the patient's needs [80].<br>(xv) Demonstrated an Electronic Health Record for Clinical Research (EHR4CR) semantic interoperability approach for bridging the clinical care and clinical research domains [97].<br>(xvi) SNOMED-CT ontologies were used to map big laboratory datasets with metadata in the form of clinical concepts [83].<br>(xvii) An online medical discussion forum where practitioners can start a topic-specific discussion and then the platform analyzes centrality measurements and semantic similarity metrics to find the most prominent practitioners in a discussion forum [84].<br>(xviii) Developed a UMLS-OWL conversion system to translate UMLS content into an OWL 2 ontology that can be queried and inferred via a SPARQL endpoint [93].<br>(xix) Researchers used SPA to detect illness and connect to the most excellent specialist. Besides, they recounted a schema representing a database query enabling doctors to pick and determine the most suitable EHR and patient data in healthcare scenarios [90].<br>(xx) The Semantic Web, blockchain, and Graph DB were combined to provide a patient-centric perspective on healthcare data in a cooperative healthcare system [94]. |
| ***Frontier technology*** | (i) A cluster-based missing value imputation (CMVI) preprocessing strategy for preparing raw data is designed to enhance the imputed data quality of a diabetes ontology graph [101].<br>(ii) Presented hypotheses on how image as ontology can be used in medical informatics and how ontology-based deep learning models can help computer vision [100].<br>(iii) Discussed a deep learning technique called the ontology-based restricted Boltzmann machine (ORBM) that can be used to gain an understanding of electronic health records (EHRs) [100].<br>(iv) Developed a Semantic Web app for question-answering using NLP where users can question about health-related information [27]. |

Table 3: Continued.

| Themes | Contributions |
| --- | --- |
| ***Regulatory Conditions*** | (i) One needs DAML + OIL to express sophisticated taxonomic knowledge, and rules should aid in the definition of dependencies between relations and use predicates of arbitrary, while metaclasses may be useful in taking advantage of current medical standards [103].<br>(ii) Described using Web 3.0-based social application for medical knowledge and communication with others and with faculty members [104].<br>(iii) The impact of Web 3.0 awareness on the academic performance of Rajiv Gandhi University of Health Sciences faculty and researchers was investigated [104].<br>(iv) Users can insert structured clinical information in the domains using SNOMED-CT terms [107].<br>(v) Demonstrated the congruence between health informatics and Semantic Web standards, obstacles in representing Semantic Web data, and barriers in using Semantic Web technology for web service [107].<br>(vi) The significant qualities of a Semantic Web language for medical ontologies were discussed [107]. |

Table 4: Summarization of the research gaps and future research avenues.

| Themes | Research gaps |
| --- | --- |
| ***E-healthcare service*** | (i) No interoperable healthcare system has yet been deployed [46].<br>(ii) Researchers have not yet looked into the policy limits of video as ontologies at an organizational level [61].<br>(iii) Prior research has focused solely on the limitations and policies of expanding an existing healthcare delivery system to directly recommend medications to users without the assistance of medical professionals [21].<br>(iv) Scholars are yet to investigate how Web 3.0 can be used to promote education through resource sharing [63].<br>(v) There is a dearth of studies on the role of existing healthcare applications in detecting patient severity levels based on the health data collected from patients [55].<br>(vi) Any prior studies did not take into account a system that can automatically determine, choose, and compose web services [3].<br>(vii) The challenges in big data connectivity into RDF, as well as privacy and security concerns, that were not addressed [2].<br>(viii) The extant literature includes only a few examples where researchers have developed a systematic clinical validation system based on the study [58].<br>(ix) The prior literature still cannot seem to distinguish ways to improve the similarity score between service parameters using statistics-based strategies and natural language processing techniques [64].<br>(x) Any previous studies on the Internet of Things domain did not consider the semantic interoperability assessment between healthcare data, services, and applications [65]. |
| ***Diseases*** | (i) No previous work had proposed an ontology for a healthcare system to efficiently store ontological data with proper evaluation criteria that meet W3C standards [20].<br>(ii) Researcher is yet to put them into practice a full-featured Integrated Mobile Information System for diabetic healthcare [76].<br>(iii) No prior work is done in expanding the set of prebuilt queries of a particular disease to handle a wide range of use cases through possible linked data evolutions [10].<br>(iv) Earlier studies did not consider mapping the triplets of one disease RDF to other existing medical services, applications, and administrations in order to conduct client assessments [73].<br>(v) There is a deficit of research on the development of intelligent user interfaces that understand the semantics of clinical data [74]. |
| ***Information management*** | (i) Knowledge gap in the current research in indexing higher-quality videos for better attribute extraction [94].<br>(ii) Indexing strategy for retrieving attributes from an audio-visual web system is yet to be addressed [81].<br>(iii) Need for a greater understanding of Semantic Web applications related to web mining to build ontologies for healthcare websites [79].<br>(iv) Prior literature addressed only modeling and annotation for a specific disease such as urinary tract infection diseases. The literature is yet to identify methods for generalizing clinical application models [85].<br>(v) No studies on the asymmetry minimization system take into account both the insurer's and the existing patient's perspectives [22].<br>(vi) The literature is yet to find ways to complete the WSMO generator from HL7 with a user interface [90]. |
| ***Frontier technology*** | (i) The literature is yet to find ways so that web applications can combine natural language processing (NLP) and domain knowledge induction in decision making and automate medical healthcare services [27].<br>(ii) The literature is yet to discover a technique to combine cloud computing, AI, and quantum physics with a platform to anticipate the chemical and pharmacological properties of small-molecule compounds for medication research and design [100]. |
| ***Regulatory conditions*** | (i) Lack of information on the Semantic Web tools before the authors moved onto the architecture of the system [103].<br>(ii) There was no emphasis on the semantic quality of available languages in any of the literature evaluation steps [104]. |

Table 5: Future research avenues in the form of research questions.

| Themes | Future research avenues |
|---|---|
| ***E-healthcare service*** | (i) What features should an interoperability framework contain in order to be considered complete [46]?<br>(ii) What technologies are required to generate video file ontologies, and what are the drawbacks of doing so [61]?<br>(iii) What approaches may healthcare organizations use to provide medical recommendations without consulting the medical practitioners directly [21]?<br>(iv) How can the healthcare industry use Web 3.0 to boost medical education [63]?<br>(v) What strategies can be applied to assess a patient's severity level based on the patient's collected health data [55]?<br>(vi) What technologies can be utilized to create web services, and how can a system automatically determine and choose the optimal web services for it [3]?<br>(vii) What kinds of security precautions should be considered while sharing information over the web [2]?<br>(viii) When it comes to adopting a Web 3.0-based clinical validation system, what technological skills and facility-related challenges do researchers face? What steps should be taken to ensure that clinical processes are validated [58]?<br>(ix) What strategies and techniques can healthcare organizations use to increase similarity scores between service parameters [64]?<br>(x) How can semantic interoperability between healthcare data, services, and applications be assessed in the context of the Internet of Things [65]? |
| ***Diseases*** | (i) Which policies and regulations may ontological systems use to comply with W3C standards [20]?<br>(ii) How can scholars expand a disease's set of queries to cover a wider range of use cases [10]?<br>(iii) What will be the most effective user interface designs for massive data networks that can interpret the semantics of clinical data [74]? |
| ***Information management*** | (i) What are the recommendations for indexing high-quality videos in Graph DB to increase attribute extraction [94]?<br>(ii) What procedures must be followed in order to extract attributes from the data that are gathered from different audio-visual web systems [81]?<br>(iii) Is it possible to improve the performance of the Web Service Modeling Ontology generator with a modified user interface [90]?<br>(iv) Will the RDF ontology be able to replace web crawlers in terms of retrieving required data from the web [80]? |
| ***Frontier technology*** | (i) How can web applications automate medical healthcare services by combining natural language processing (NLP) with domain knowledge induction in decision making [27]?<br>(ii) How could the Semantic Web platform anticipate the chemical and pharmacological properties of small-molecule compounds using cloud computing, quantum physics, and artificial intelligence [100]?<br>(iii) What are the procedures to implement ontology-based restricted Boltzmann machine (ORBM) in electronic healthcare record (EHR) [100]? |
| ***Regulatory conditions*** | (i) Which techniques can be used to optimize NLP for transforming pathology report segments into XML [103]?<br>(ii) What strategies and activities may developers employ to address semantic quality issues in existing languages [104]? |

Rather than going to the hospital for help and support, patients' health-related parameters can now be monitored remotely, constantly, consistently, and in real time and then processed and transferred to medical data enters via cloud storage. Because of cloud platforms' security risks, choosing one is a major technological challenge for the healthcare industry. Some of these cloud-based storage systems cannot adequately preserve patients' data and information regarding semantic data [6, 8]. However, none of the research articles suggested any architectures, nor were any intended prototypes built to address these cloud security issues of MIoT in general.

*4.5. Scope of Regulatory Condition Research.* Regulations are paramount for the healthcare and medical industries to function properly. They support the global healthcare market, ensure the delivery of healthcare services, and safeguard patients,' doctors,' developers,' researchers,' and healthcare agents' rights and safety. The Semantic Web also has its detractors, like many other technologies, in terms of legislation and regulation. Historically, scaling medical knowledge graphs has always been a challenge. As a result of privacy and legal clarity, healthcare companies are not sufficiently incentivized to share their data as linked data. Only a few academic papers and documents disclose how these corporations use to automate the process. Furthermore, compared to other types of datasets, many linked datasets representing tools are of poor quality. As a result, applying them to real-world problems is highly challenging. Other alternatives, such as property graph databases like Neo4j and mixed models like OriendDB, have grown in popularity due to the RDF format's complexity. Healthcare application developers and designers prefer to use web APIs over SPARQL endpoint to send data in JSON format. This study illustrates that more research is needed to improve the semantic quality of available technologies (e.g., RDF, OWL, and SPARQL) to effectively use them in the healthcare industry to ease healthcare development.

## 5. Discussion

This section describes the findings from the selected studies based on answer to the research questions. Therefore, the readers will be able to map the research questions with the contribution of this systematic review.

*5.1. (RQ1) What Is the Research Profile of Existing Literature on the Semantic Web in the Healthcare Context?* The research aims to determine the primary objectives of using the Semantic Web and the major medical and healthcare sectors where Semantic Web technologies are adopted. As the Semantic Web has shown incremental research trends in recent years, there is a need for a structured bibliometric study. This study collected data from the Scopus, IEEE Xplore Digital Library, ACM Digital Library, and Semantic Scholar databases, focusing on various aspects and seeing their affinity. We performed bibliometric analysis to look at essential details like preliminary information, country, author, and application area where these publications are being used for the Semantic Web in the context of healthcare. We conducted the bibliometric analysis using an open-source application called VOS viewer. The outcomes and specifics of the experiment are detailed in Section 2.

As stated in the methodology section, our study consists of 65 documents. A number of prestigious conferences, publications, and events have published these healthcare-related articles. Out of these 65 shortlisted papers, 27 were presented in conferences, 21 in journals, and 17 from book chapters. Our study observes that the field of "Semantic Web in Healthcare" is not comparatively new. The first paper from the shortlisted documents on this topic was published in 2001. Since then, there has been minimal growth in this field, with 2007 appearing to be the start. Surprisingly, the maximum number of articles (8) published in this discipline was in 2013, but from 2013 to 2016, there was only a minor shift by researchers globally. It is most likely due to the introduction of Web 3.0 in 2014. It is yet to be found how Web 3.0 will effectively leverage the Semantic Web as a core component rather than seeing it as a competing technology in the medical healthcare field. The decrease in the number of articles shows how the interests of researchers switched from the Semantic Web to the emerging Web 3.0. However, the Semantic Web remains the top choice of medical practitioners as Web 3.0 evolves. Furthermore, the United States is the country with the most research papers, followed by France and India (see Figure 4). It implies that both developed and emerging countries use the Semantic Web in their healthcare industries. VOS viewer also discovered 35 works titled to be published in Computer Science, 16 in Engineering, 9 in Medicine, and 5 in Mathematics. We also used the VOS viewer software to visually represent the keyword co-occurrences from those shortlisted 65 publications. The total number of keywords was 774. The minimum number of times a keyword appears is set at 5. The terms that occurred more than five times in all texts are included in our representation. We found 76 keywords that meet our requirements. Figure 7 shows our findings in a co-occurrence graph containing the other essential phrases. As expected, Semantic Web and healthcare are the most occurring keywords, and both are mentioned 55 times. Following that, web services, decision support systems, interoperability, etc. are listed. These terms are used to categorize the Semantic Web's application areas in healthcare.

Our analysis also reveals that most proposed frameworks for improving and expanding the healthcare system do so without the involvement of health professionals. Some of them discussed data interoperability, diseases, frontier technologies, and regulatory issues, while others emphasized the use of video as ontologies and video conferences in bridging communication gaps. The majority of the publications only propose frameworks with no implementation. Web services currently merely make services available, with no automatic mechanism to connect them in a meaningful way.

*5.2. (RQ2) What Are the Primary Objectives of Using the Semantic Web, and What Are the Major Areas of Medical and Healthcare Where Semantic Web Technologies Are Adopted?* The adoption of the Semantic Web in healthcare strives to improve collaboration, research, development, and organizational innovation. The Semantic Web has two primary objectives: (1) facilitating semantic interoperability and (2) providing end-users with more intelligent support. Semantic interoperability, a key bottleneck in many healthcare applications, is one of today's major problems. Semantic Web technologies can help with data integration, knowledge administration, exchange of information, and semantic interoperability between healthcare information systems. It focuses on building a web of data and making it appropriate for machine processing with little to no human participation. So, healthcare computer programs can better assist in finding information, personalizing healthcare information, selecting information sources, collaborating within and across organizational boundaries, and so on by inferring the consequences of data on the Internet.

Based on our review of the findings, we found five application domains where the Semantic Web is being adopted in the healthcare context. This study will brief those domains from Sections 5.2.1 to 5.2.5 as well as justify them in relation to healthcare.

*5.2.1. E-Healthcare Service.* More than two-fifth of the total studies (65) considered in this study is about e-healthcare services (see Table 2). These studies focus on ways to use the Internet and related technologies to offer and promote health services and information, as well as diagnosis recommendation systems and online healthcare service automation.

In this study, researchers developed a web-based prototype that generates the required reports with a high degree of data integration and a rule-based production technique for establishing a link between prevalent diseases and the range of diseases in a specific gene [64].

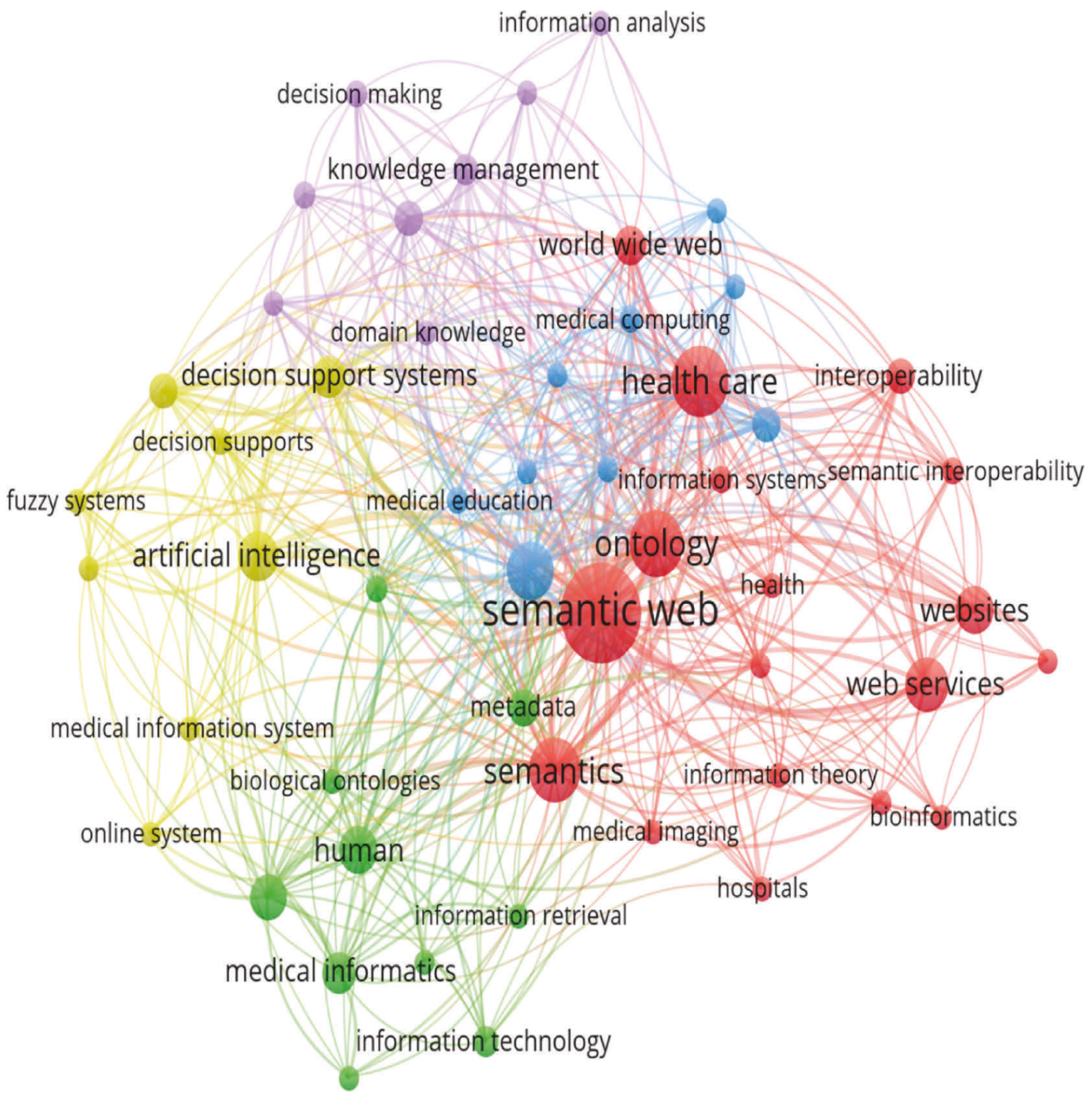

Figure 7: Co-occurrence network of the index's keywords.

Another group of e-healthcare service studies focused on how current electronic information and communication technology could help people's health and healthcare [46, 49, 50, 61–64, 97]. Most of the authors used a WSMO (Web Service Modeling Ontology) service delivery platform and an automatic alignment of user-defined EHR (electronic health record) workflows, where service owners can register a service, and the system will automate prefiltering, discovery, composition, ranking, and invocation of that service to provide healthcare.

The adoption of e-healthcare in developing countries has shown to be a feasible and effective option for improving healthcare. It allows easy access to health records and information and reduces paperwork, duplicate charges, and other healthcare costs. If the proper implementation of e-healthcare technologies is ensured, everyone will benefit.

*5.2.2. Diseases.* Out of 65 articles, there are only 8 articles regarding the adoption of the Semantic Web in the diseases sector (see Table 2). These articles present a discussion on the deployment of a disease-specific healthcare platform, disease information exchange system, knowledge base generation, and research portal for a specialized disease.

This study developed a web-based prototype for an Integrated Mobile Information System (IMIS) for diabetic patient care [20]. The authors used ontology mapping so that related organizations could access each other's information. They also embedded feedback and communication mechanisms within the system to include user feedback.

Another study developed queryMed packages for pharmaco-epidemiologists to access and link medical and pharmacological knowledge to electronic health records [10]. The authors distinguished all the medications endorsed for critical limb ischemia (CLI) and recognized one contraindicated solution for one patient.

Disease management/prediction systems are necessary for finding the hidden knowledge within a group of disease data and can be used to analyze and predict the future behavior of diseases. An all-in-one strategy rarely works in the healthcare industry. It is critical to develop a personalized and contextualized disease prediction system to enhance user experience.

*5.2.3. Information Management.* Almost two-fifths of the total studies considered in this study (65) are about information management (see Table 2). After e-healthcare service, this category has the most studies. These articles are particularly about healthcare management systems, medical information indexing, healthcare interoperability systems, decision making, coordination, control, analysis, and visualization of healthcare information.

This study presented a medical knowledge morphing system that focuses on ontology-based knowledge articulation and morphing of heterogeneous information using logic and ontology mediation [105]. The authors used high-level domain ontology to describe fundamental medical concepts and low-level artifact ontology to capture the content and structure.

In another study, an annotation image (AIM) ontology was developed to provide important semantic information within photographs, allowing radiological images to be mined for image patterns that predict the structures' biological features. The authors transformed XML data into OWL and DICOM-SR to control ontological terminology in order to create image annotation.

A well-designed healthcare information system is required for management, evaluation, observations, and overall quality assurance and improvement of key stakeholders of the health system. Even though a significant amount of work is done in this sector, it is far from sufficient. It is something on which we should focus.

*5.2.4. Frontier Technology.* We found only 3 publications on frontier technology (see Table 2). These articles describe healthcare application domains that use AI, machine learning, or computer vision to automate medical coding, generate medical informatics, and deal with intelligent IoT data and services.

The first review article is about a method for preprocessing raw cluster-based missing value imputation (CMVI), with the goal of improving the imputed data quality of a diabetes ontology graph [27]. Their findings show that preprocessed data have better imputation accuracy than raw, unprocessed data, as measured against coefficient of determination (R2), index of agreement (D2), and root mean square error (RMSE).

Another article talks about ideas on how image as ontology can be used in health informatics and how deep learning models built on ontologies can support computer vision [100].

Frontier technology such as AI, ML, and IoT offers many advantages over traditional analytics and clinical decision-making methodologies. At a granular level, those technologies provide

(i) Increased efficiency.
(ii) Better treatment alternatives.
(iii) Faster diagnosis.
(iv) Faster drug discovery.
(v) Better disease outbreak prediction.
(vi) Medical consultations with patients with little or no participation of healthcare providers.

There is a lack of research on the integration of frontier technologies with the Semantic Web. Researchers should focus their efforts on this area. Students must take the initiative to develop creative technological inventions.

*5.2.5. Regulatory Conditions.* There were only 3 publications that used Semantic Web technology to address regulatory conditions (see Table 2). These studies focus on the challenges and requirements of the Semantic Web and technologies that represent the Semantic Web, awareness, and policy and regulations.

An article describes how to design, operate, and extend a Semantic Web-based ontology for an information system of pathology [103]. The authors of this paper highlight what technologies, regulations, and best practices should be followed during the entire lung pathology knowledge base creation process.

Another study talks about the challenges of integrating healthcare web service composition with domain ontology to implement diverse business solutions to accomplish complex business logic [104].

Privacy and regulation are important in establishing a clear framework within which healthcare providers, patients, healthcare agents, and healthcare application developers can learn and maintain the skills needed to provide high-quality health services which are safe, productive, and patient-centered. From these regulatory condition-type articles, we can understand whether technology is easy to use, has challenges, and is emerging, secure, and valuable to the healthcare community. We need to do more work on this.

*5.3. (RQ3) Which Semantic Web Technologies Are Used in the Literature, and What Are the Familiar Technologies Considered by Each Solution?* This section discusses the various Semantic Web technologies used in the literature, as well as the most common ones among them. There are numerous Semantic Web technologies available that make the applications more advanced. The healthcare industry makes extensive use of these Semantic Web technologies. As a result of these technologies, the healthcare industry is getting more advanced. The most prevalent Semantic Web technologies that are used in the healthcare sector are Resource Description Framework (RDF), Web Ontology Language (OWL), SPARQL Protocol and RDF Query Language (SPARQL), Semantic Web Rule Language (SWRL), Web Service Modeling Ontology (WSMO), Notation3 (N3), SPARQL Inferencing Notation (SPIN), Euler Yap Engine (EYE), Web Service Modeling Language (WSML), and RDF Data Query Language (RDQL).

Various Semantic Web technologies are used to accomplish various goals, such as converting relational databases to RDF/OWL-based databases, data linking, reasoning, data sharing, data representation, and so on. Ontologies are considered the basis of the Semantic Web. All of the data on the Semantic Web are based on ontologies. To take advantage of ontology-based data, it must first be transformed into RDF-based datasets. The RDF is an Internet standard model for data transfer that includes qualities that make data merging easier, as well as the ability to evolve schemas over time without having to update all of the data [52]. The majority of the researchers utilized RDF to represent the linked data and interchange data. In the Semantic Web, Notation3 is used as an alternative to RDF to construct notations. It was created to serialize RDF models and it supports RDF-based principles and constraints. Humans can understand Notation3-based notations more easily than RDF-based notations. In addition to RDF, OWL is employed in the research articles to express ontology-based data. The OWL is a semantic markup language for exchanging and distributing ontologies on the web [52]. Furthermore, there is a second version of OWL available which is known as OWL2. The improved descriptive ability for attributes, enhanced compatibility for object types, simplified metamodeling abilities, and enhanced annotation functionality are among the new features added in OWL2. Numerous OWL-based ontologies are available on the web. OWL-S is one of them which is a Semantic Web ontology [78]. The OWL is also used for semantic reasoning. Combining Description Logic with OWL (OWL-DL) takes the reasoning capability to another level. OWL-DL provides desired algorithmic features for reasoning engines and is meant to assist the current Description Logic industry area [82]. As an alternative to OWL, EYE is used which is an advanced chaining reasoner with Euler path detection [85]. It uses backward and forward reasoning to arrive at more accurate conclusions and results. To query the RDF and OWL-based datasets, the scholars made use of SPARQL. SPARQL is the sole query language that may be used to query RDF and OWL-based databases. However, RDQL was employed as a query language for RDF datasets in a study [20]. Only RDF datasets can be queried with it. In several papers, writing the semantic rules and constraints was necessary. So, they used SWRL which is a language for writing semantic rules based on OWL principles. Alongside SWRL, scholars used SPIN which is a rule language for the Semantic Web that is based on SPARQL [60]. In the Semantic Web, specifying web services for different purposes is essential. In this regard, some research papers discussed leveraging the WSMO which is a Semantic Web framework for characterizing and specifying web services in a semantic way. A linguistic framework called WSML is used to express the Semantic Web services specified in WSMO. The WSML is a syntactic and semantic language framework for describing the elements in WSMO [48]. Tables 4–8 summarize the Semantic Web technologies employed in different thematic research areas. Section 3 has detailed information regarding the discussion.

Table 6 summarizes Semantic Web technologies used in e-healthcare services. In this field of theme research, RDF, OWL, and SPARQL are the most commonly utilized technologies. Researchers employed RDF and OWL to construct RDF-based datasets, represent RDF datasets, and develop links between data. As an alternative to RDF, an article used Notation3 to construct RDF notations which are easier to read than RDF-based notations. In a paper, the scholars used OWL2, the second version of OWL, to utilize the latest features offered by the technology. For all of the articles, SPARQL is the only query language utilized to query the datasets. To construct rules and limits for the systems, most of the articles used SWRL. In addition to SWRL, an article used the SPIN to generate semantic rules and constraints. Furthermore, SPIN has not been used in any other research area. Besides, two articles used WSMO for the identification of Semantic Web services required for the systems. On the other hand, three articles in this theme did not use any Semantic Web technology.

Table 7 summarizes Semantic Web technologies used in diseases. Similar to the preceding thematic research area, RDF, OWL, and SPARQL are the most frequently used technologies. Also, the motivations for using these technologies are identical. However, an article utilized RDQL as an alternative to SPARQL to conduct queries on RDF datasets. SWRL was used to construct rules and limitations, just as it had been previously. It is also worth noting that a study built a model using the OWL-S, an OWL-based semantic ontology. Then, there is a study in this field that did not utilize any Semantic Web technology at all.

Table 8 summarizes Semantic Web technologies used in information management. Nine distinct Semantic Web technologies are used in this thematic research area. RDF, OWL, and SPARQL, like the previous topic groups, are the most extensively used technologies. It is worth repeating that the technologies' goals are the same as they were previously. In addition, the usage of Notation3 for more accessible RDF notations, OWL2 to take advantage of new capabilities, OWL-S semantic ontology as the data source, and WSMO to identify Semantic Web services are also mentioned in this thematic area. In this field of research, there are two new technologies that are not present in prior fields. OWL-DL, which combines OWL with Description Logic for information reasoning, is one of the new technologies. The other one is EYE reasoner, which is also a reasoning engine. On the contrary, a significant proportion of articles, six to be exact, did not employ any Semantic Web technologies.

Table 7 summarizes Semantic Web technologies used in frontier technology. In this thematic study field, there are just three articles, and two of them did not employ any kind of Semantic Web technology. The other paper includes RDF and SPARQL, which were very commonly used in the prior thematic research fields.

Table 8 summarizes Semantic Web technologies used in regulatory conditions. Only one of the two articles in this research area includes Semantic Web technology. Also, the sole semantic technology used in the article is RDF for the purpose of semantic data representation.

There are different applications of Semantic Web technologies in the articles, but most of the technologies are common in several articles. The most commonly used Semantic Web technologies are the SPARQL query language, RDF, OWL, and SWRL. Almost 80 percent of the analyzed papers used different functionalities of RDF. Furthermore, OWL and SPARQL technologies were used in nearly three-quarters of the articles. Besides, SWRL technology was applied in one-third of the analyzed studies. It is now obvious that these technologies have the potential to improve the healthcare industry.

*5.4. (RQ4) What Are the Evaluating Procedures Used to Assess the Efficiency of Each Solution?* The suggested technologies and procedures for evaluating these works are included in this category. In truth, assessing the designed healthcare system's quality, performance, and utility is a crucial responsibility. Because the healthcare industry is highly sensitive, suitable evaluation standards are necessary. Due to technological limitations, however, the evaluation system is not well organized or maintained. Because the notion of Semantic Web technology is new in the medical field, overall development and evaluation are inadequate.

Table 6: Summary of Semantic Web technologies used in e-healthcare services.

| References | RQ3 | RQ4 |
|---|---|---|
| [2] | SPARQL | × |
| [3] | × | × |
| [4] | RDF, OWL2, SPARQL, WSMO | × |
| [17] | OWL, SPARQL | × |
| [21] | OWL | × |
| [26] | RDF, SPARQL | × |
| [48] | OWL, WSML | × |
| [47] | RDF, OWL | × |
| [46] | OWL | × |
| [49] | RDF, OWL, SPARQL | × |
| [50] | RDF, SPARQL | × |
| [51] | WSMO | Flora-2 Expression |
| [53] | RDF, OWL, SPIN | × |
| [52] | RDF, OWL, SPARQL, SWRL, Notation3 | × |
| [55] | RDF, OWL, SWRL | × |
| [54] | RDF, OWL, Notation3 | × |
| [57] | RDF, OWL, SPARQL, SWRL | OSHCO Validation |
| [56] | OWL, SWRL | × |
| [58] | OWL, SPARQL, SWRL, | Histopathology Method |
| [60] | RDF, OWL, SPARQL, SWRL, SPIN | × |
| [62] | RDF, OWL, SPARQL, SWRL | WS Composition System |
| [61] | × | × |
| [63] | × | × |
| [64] | SPARQL | × |
| [65] | RDF | × |
| [66] | RDF, OWL | × |
| [67] | RDF, OWL, SPARQL | × |

Table 7: Summary of Semantic Web technologies used in diseases.

| References | RQ3 | RQ4 |
|---|---|---|
| [10] | RDF, OWL, SPARQL | × |
| [20] | RDF, OWL, RDQL | × |
| [71] | RDF, OWL, SPARQL | × |
| [72] | RDF, OWL, SWRL | × |
| [73] | × | × |
| [74] | RDF, OWL, SPARQL | × |
| [76] | × | × |
| [78] | RDF, OWL, OWL-S | × |

In the e-healthcare service-based theme (see Table 6), the authors in [51] established a set of setups to test the matcher's efficiency for scalability in terms of the number of Semantic Web services for medical appointments and their complexity. They consider the logical complexity of Flora-2 expressions used in pre and post-conditions, which can handle various web service and goal descriptions, including ontology consistency check. Some other evaluation methods like OSHCO validation for automatic decision support in medical services were also introduced by the authors in [57]. An experiment was established to assess the system utilizing two metrics via WS datasets, the execution time measurement and the correctness measurement, for graph-based Semantic Web services for healthcare data integration [62] and histopathology for evaluating the performance of semantic mappings [58].

However, only two publications presented evaluation procedures from the vast portion of information management system-related work (see Table 7). Tonguo et al. [25] used BioMedLib to evaluate a system that takes a user's search query and pulls articles from millions of national biomedical article databases. Another one used evaluation criteria like D2RQ for default semantic mapping generation [83].

In terms of frontier technology (see Table 9), the cluster-based missing value imputation algorithm (CMVI) was used to extract knowledge in the Semantic Web's healthcare domain [101]. The imputation accuracy was measured using a couple of well-known performance metrics, namely, coefficient of determination (R2) and index of agreement (DK), along with the root mean square error (RMSE) test. In addition, various open-domain question-answer evaluation campaigns such as TREC21, CLEF22, NTCIR23, and Quaero24 have been launched to evaluate a Semantic Web and NLP-based medical questionnaire system [27].

None of the writers provide any evaluation methodologies connected to diseases and regulatory conditions (see Tables 9 and 10). To assess the consequences of Semantic Web discussions on specific diseases, well-designed evaluation criteria are required. As studies focus on the obstacles and problems of the Semantic Web in healthcare services, the necessity of evaluation is also missing in regulatory conditions.

Table 8: Summary of Semantic Web technologies used in information management.

| References | RQ3 | RQ4 |
|---|---|---|
| [1] | × | × |
| [22] | RDF, OWL | × |
| [25] | RDF, OWL, SPARQL | BioMedLib (Deployment Model) |
| [79] | RDF, OWL | × |
| [80] | OWL, SWRL | × |
| [70] | RDF, SPARQL | × |
| [81] | × | × |
| [82] | OWL, OWL-DL | × |
| [83] | RDF, SPARQL | D2RQ Framework |
| [84] | RDF, SPARQL | × |
| [85] | RDF, Notation3, EYE | × |
| [86] | RDF, OWL, SPARQL, SWRL | × |
| [87] | RDF, OWL | × |
| [88] | × | × |
| [89] | × | × |
| [90] | OWL-S, WSMO | × |
| [91] | RDF, OWL-S, WSMO | × |
| [92] | × | × |
| [93] | OWL, OWL2, SPARQL | × |
| [110] | RDF, OWL, SPARQL | × |
| [94] | × | × |
| [106] | RDF, OWL, SPARQL | × |
| [97] | RDF, SPARQL | × |
| [105] | × | Stovanojic's Ontology Evolution and Management Process |

Table 9: Summary of Semantic Web technologies used in frontier technology.

| References | RQ3 | RQ4 |
|---|---|---|
| [27] | RDF, SPARQL | QA Evaluation (TREC, CLEF, NTCIR, Quaero) |
| [100] | × | × |
| [101] | × | Root mean square error (RMSE) |

*5.5. (RQ5) What Are the Research Gaps and Limitations of the Prior Literature, and What Future Research Avenues Can Be Derived to Advance Web 3.0 or Semantic Web Technology in Medical and Healthcare?* The healthcare industry is on the verge of a real Internet revolution. It intends to bring in a new era of web interaction through the adoption of the Semantic Web, with significant changes in how developers and content creators use it. This web will make healthcare web services, applications, and healthcare agents more intelligent and even provide care with human-like intelligence by utilizing an AI system. Despite the tremendous amount of innovation, it may bring its adoption in healthcare considerable challenges.

The problem with the “Semantic Web” is that it requires a certain level of implementation commitment from web developers and content creators that will not be forthcoming. First, a large portion of existing healthcare web content does not use semantic markup and will never do so due to a lack of resources to rewrite the HTML code. Second, there is no guarantee that new healthcare content will utilize semantic markup because it would need additional effort.

Table 10: Summary of Semantic Web technologies used in regulatory conditions.

| References | RQ3 | RQ4 |
|---|---|---|
| [103] | RDF | × |
| [104] | × | × |
| [107] | OWL | × |

However, it is essential to guide the Semantic Web developer community in the right direction so that they can help contribute to future medical healthcare development. The following are the primary obstacles the Semantic Web faces in general: (i) content availability, (ii) expanding ontologies, (iii) scalability, (iv) multilingualism, (v) visualization to decrease information overload, and (vi) Semantic Web language stability.

Furthermore, based on our thorough examination of the 65 publications, the following are some of the most technologically severe obstacles that the Semantic Web in general faces in the healthcare context and must overcome; future research may be able to alleviate a few of these challenges:

(1) *Integrated Data Issue*. The vulnerability of interconnected data is one of the most significant challenges with Semantic Web adoption. All of a patient's health records and personal information are stored and interlinked to an endpoint, and a malicious party may gain control of one's life if the record is compromised.

(2) *Vastness*. The current Internet contains a vast amount of healthcare records not yet semantically indexed; any reasoning system that wants to analyze all of these data and figure out how it functions will have to handle massive amounts of data.

(3) *Vagueness*. As Semantic Web is not yet mature enough, applications cannot handle non-specific user queries adequately.

(4) *Accessibility*. Semantic Web may not work on older or low-end devices; only highly configured devices will be able to manage web content.

(5) *Usability*. It will be difficult for beginners to comprehend because the SPARQL queries are often used in websites and services.

(6) *Deceit*. What if the information provided by the source is false and deceptive? Management and regulation have become crucial.

The study also identifies future research opportunities and gives research recommendations to the developer and researcher communities for each of the identified theme areas where the Semantic Web is being used in medical and healthcare (see Section 4). Tables 4 and 5 summarize the research gap and probable future research direction.

## 6. Conclusion

The purpose of this SLR is to discover the most recent advances in SW technology in the medical and healthcare fields. We used well-established research techniques to find relevant studies in prestigious databases such as Scopus, IEEE Xplore Digital Library, ACM Digital Library, and Semantic Scholar. Consequently, we were able to answer five significant RQs. We answered **RQ1** by giving a bibliometric analysis-based research profile of the existing literature. The study profile includes information on annual trends, publishing sources, methodological approaches, geographic coverage, and theories applied (see Sections 2.4 and 5.1). We performed content analysis to determine the answers to **RQ2**, **RQ3**, and **RQ4**; we also identified research themes, with a focus on technical challenges in healthcare where SW technologies can be used (see Sections 3 and 5.2–5.4). Finally, the synthesis of prior literature helped us to identify research gaps in the existing literature and suggest areas for future research in **RQ5** (see Section 5.5 and Tables 4 and 5). The findings of this study have important implications for healthcare practitioners and scholars who are interested in the Semantic Web and how it might be used in medical and healthcare contexts.

The global digital healthcare market is growing to meet the health needs of society, individuals, and the environment. As a result, a substantial study is required to assist governments and organizations in overcoming technological challenges. We successfully reviewed 65 academic papers comprising journal articles, conference papers, and book chapters from prestigious databases. We have identified five thematic areas based on our research questions to discuss the objectives, solutions, and prior work of Semantic Web technology in the healthcare field. Among these, we observed that e-healthcare services and medical information management are the most discussed topics [105, 107]. According to our findings, with the emergence of Semantic Web technology, integration, discovery, and exploration of medical data from disparate sources have become more accessible. Accordingly, medical applications are incorporating semantic technology to establish a unified healthcare system to facilitate the retrieval of information and link data from multiple sources. Most of the studies that we examined discussed the importance of knowledge sharing among clinicians and patients to develop an effective medical service. The frameworks described depended on the proper data distribution from various sources supported by specific technology interventions [24]. To answer patient queries, SW-based systems such as appointment matchmaking, quality assurance, and NLP-based chatbots have been proposed to improve healthcare services [24, 111, 112]. In short, the Semantic Web has huge potential and is widely regarded as the web's future, Web 3.0, which will present a new challenge and opportunity in combining healthcare big data with the web to make it more intelligent [6, 113].

The analysis of the proposed solutions discussed in the papers helped us to identify the main challenges in healthcare systems. Besides that, this study also identifies future challenges and research opportunities for future medical researchers. We observed that most of the proposed solutions are yet to be implemented and many problems are only rudimentarily tackled so far. In conclusion, by exchanging knowledge among physicians, researchers, and healthcare professionals, the SW encourages improvement from the "syntactic" to "semantic" and finally to the "pragmatic" level of services, applications, and people. From the overall observation of the findings of this SLR, a future strategy will be to adopt some of the suggested solutions to overcome the shortcomings and open a new door for the medical industry. In the future, we will try to implement such solutions and eliminate the problems.

## Data Availability

The data used to support the findings of this study are included within the article.

## Conflicts of Interest

The authors declare that they have no conflicts of interest.